\documentclass[5p]{elsarticle}
\usepackage{graphicx}
\usepackage{caption}
\usepackage{subcaption}
\usepackage{hyperref}
\usepackage[T1]{fontenc}
\usepackage{geometry}
\usepackage{mathtools}  

\usepackage{multirow,color}
\usepackage[table]{xcolor}
\usepackage{algpseudocode}
\usepackage{tikz}
\usepackage{nccmath}
\usepackage[linesnumbered,ruled,vlined]{algorithm2e}
\usepackage{float}
\usepackage[font=small,labelfont=bf]{caption}
\usepackage{mathtools, cuted}
\usepackage{lipsum, color}

\RequirePackage{fix-cm}

\usepackage{graphicx}
\usepackage{caption}
\usepackage{subcaption}
\usepackage{hyperref}
\usepackage[T1]{fontenc}
\usepackage{geometry}
\usepackage{mathtools}  
\usepackage{multirow,color}
\usepackage[table]{xcolor}
\usepackage{algpseudocode}
\usepackage{tikz}
\usepackage[linesnumbered,ruled,vlined]{algorithm2e}
\usepackage{float}
\usepackage[font=small,labelfont=bf]{caption}
\usepackage{mathtools, cuted}
\usepackage{lipsum, color}
\usepackage{ marvosym }

\usepackage{microtype}
\usepackage{indentfirst}
\usepackage{relsize}
\usepackage{makecell}

\journal{arXiv}
\begin{document}
\begin{frontmatter}

\title{Semi-fragile Tamper Detection and Recovery based on Region Categorization and Two-Sided Circular Block Dependency}

\author[ourmainaddress]{Seyyed Hossein Soleymani}
\cortext[mycorrespondingauthor]{Corresponding author}
\ead{seyyedhosein.soleymani@stu.um.ac.ir}

\author[ourmainaddress]{Amir Hossein Taherinia\corref{mycorrespondingauthor}}
\ead{taherinia@um.ac.ir}

\address[ourmainaddress]{Department of Computer Engineering, Faculty of engineering, Ferdowsi University of Mashhad, Mashhad, Iran}

\begin{abstract}
This paper presents a new semi-fragile algorithm for image tamper detection and recovery, which is based on region attention and two-sided circular block dependency. This method categorizes the image blocks into three categories according to their texture. In this method, less information is extracted from areas with the smooth texture, and more information is extracted from areas with the rough texture. Also, the extracted information for each type of blocks is embedded in another block with the same type. So, changes in the smooth areas are invisible to Human Visual System. To increase the localization power a two-sided circular block dependency is proposed, which is able to distinguish partially destroyed blocks. Pairwise block dependency and circular block dependency, which are common methods in the block-based tamper detection, are not able to distinguish the partially destroyed blocks. Cubic interpolation is used in order to decrease the blocking effects in the recovery phase. The results of the proposed method for regions with different texture show that the proposed method is superior to non-region-attention based methods.
\end{abstract}


\end{frontmatter}

\section{Introduction}
Watermarking can be categorized into fragile, semi-fragile and robust watermarking~\cite{cox2007digital}. In fragile watermarking, the embedded watermark will be destroyed after both of intentional attacks (such as image cropping, image copy-move forgery, and other image tampering operations) and unintentional attacks (such as image compression and image enhancement operations)~\cite{nguyen2016reversible, yu2015new, ghosal2014binomial, chen2014chaos,qin2016fragile}. So,  this type of watermarking is suitable for authentication of image. Robust watermarking is robust against both of intentional and unintentional attacks~\cite{taherinia2010new, soleymani2015robust, soleymani2016double}. So, this type of watermarking is suitable for copyright protection. Finally, semi-fragile watermarking is robust against unintentional attacks and also is fragile against intentional attacks and it reveals the tampered locations ~\cite{preda2013semi, shi2016region, huo2014semi}. Given that on the Internet, operations such as image compression, image quality enhancement, communication noise and image tampering are common, so it is need to semi-fragile watermarking for authentication and recovery of the tampered image. For recovery of the tampered image, an image digest must be created and embedded as a watermark in the original image. The watermark embedding and tamper localizaion must be imperceptible and precise, respectively, and finally, the tampered regions must be recovered with high quality.

So far, many fragile watermarking methods have been proposed for tamper detection and recovery but the number of semi-fragile watermarking methods are not so much, because there are some constraints in semi-fragile watermarking such as the limitation of robust locations for embedding. The structure of this article is as follows. The related works and the proposed method are described in sections~\ref{lbl.relatedworks} and ~\ref{lbl.proposedmethod}, respectively. Also, the experimental results and the conclusions are described in sections~\ref{lbl.experimentalresults} and~\ref{lbl.conclusions}, respectively.

\section{Related Works}\label{lbl.relatedworks}

In this section, state-of-the-art methods are reviewed that focus on the semi-fragile image watermarking for tamper detection and recovery.

In~\cite{chamlawi2010digital, ullah2013dual}, two semi-fragile image watermarking methods for tamper detection and recovery are proposed, which the main embedding algorithm of them are similar. In~\cite{chamlawi2010digital}, two watermarks are created for tamper detection and recovery, separately. But in~\cite{ullah2013dual}, just one watermark is created for both purposes. Therefore, the watermarked image quality has improved because the size of watermark is smaller and the amount of change in the original image is less. In~\cite{chamlawi2010digital}, the authentication watermark is created using a key randomly and the recovery watermark is created using some low frequency coefficients of DCT transform of the original image. These two created watermarks are compressed using Haffman codding and then to increase the robustness the BCH error correction coding is applied on the compressed watemarks. In both of~\cite{chamlawi2010digital} and~\cite{ullah2013dual} metods, the watermark information is embedded in detail sub-bands of Integer Wavelet Transform (IWT) of the original image. The group quantization of coefficients is used as the embedding method. The robustness and quality of the recovered image are not high in these two methods.

In~\cite{wang2011novel, wang2014novel}, two methods are proposed that the foundation of them are similar. In other words, method~\cite{wang2014novel} is an enhancement on the quality of method~\cite{wang2011novel}. In~\cite{wang2011novel}, six bits are created randomly as the authentication watermark for each $8\times8$ block. Each $8\times8$ block is divided into four $4\times4$ sub-blocks, and the average values of gray levels of sub-blocks are used as recovery watermark. The authentication watermark is embedded into the low-frequency coefficients of DCT transform of a paired block, which the paired block is in a different place. The four average values of recovery watermark are embedded in the middle frequency coefficients of DCT transform of a paired block. Embedding method for authentication watermark is quantization and for recovery watermark is substitution in the coefficients. Novelty of method~\cite{wang2011novel} is in recovery phase. The four average values are extracted and modified using linear regression in order to make the DC0 and three low-frequency coefficients of an $8\times8$ zero blocks. After that, the inverse DCT is applied on the created $8\times8$ block. The robustness and quality of method~\cite{wang2011novel} is high but it has the blocking effects. The mothod~\cite{wang2014novel} has tried to solve the blocking effects using a linear optimization mechanism and the estimation of the lost coefficients in a DCT transform. The results of mothod~\cite{wang2014novel} are much better than the results of mothod~\cite{wang2011novel}.

In~\cite{li2015semi}, a block-based method is proposed, which the image is divided into $16\times16$  non-overlapping blocks and the average value of each block is calculated. Five most significant bits (5-MSB) of each average value are used as the recovery watermark for each block. In this method one-sided circular block dependency is used in the embedding and tamper localization phases. Look back to the previous block in the circular dependency is used for increasing the localization power. The detail sub-bands coefficients of the second level of IWT transform are used as the place of embedding.  The quantization of maximum value in a group of coefficients is used as the embedding method.
The quality of recovered regions for the tampered image is about 20 dB based on the PSNR measure. The localization power is reported near to zero based on the false rejection (FR) and false acceptance (FA) measures. This method has a high robustness against to unintentional attacks.

In \cite{li2014semi}, a method is proposed that is based on compressive sensing. The compressive sensing is used for estimation of the missed coefficents. In this method, low frequency coefficients of $4\times4$ blocks are embedded into LH1 and HL1 sub-band of DWT transform using the substitution embedding method. Robustness and the quality of the recovered image are acceptable.

In \cite{korus2014iterative}, a method is proposed that is based on random sampling and image inpainting. In this method, some lines of pixels are selected in random directions and then the DCT transform of each line is calculated.  Some low-frequency coefficients are used as recovery watermark and are embedded in the middle-frequency coefficients of blocking DCT using the quantization based embedding method. In this method, inpainting is used for each pixel or region that there is no information about it. Robustness of this method is high but the quality of the recovered image is not high.

In \cite{phadikar2012novel}, a method is proposed in which the halftone image of two-level IWT is used as recovery watermark. The authentication watermark is created using a key randomly. The recovery watermark is embedded in LH1 and HL1 sub-bands and the authentication watermark is embedded in LH2 and HL2 sub-bands, respectively using dither-QIM embedding method. Quality of recovered image is not high because the halftone image is created using two-level IWT and a Gaussian kernel convolution is used for calculation of inverse halftoning.

In~\cite{benrhouma2015tamper}, the average value of each block is calculated and then it is normalized using some calculations to be suitable for substitution embedding method in detail sub-bands of DWT transform. The novelty of this method is embedding the decimal values using the proposed calculations.

In~\cite{lin2011roi, li2016semi, li2016semi}, three methods are proposed for the protection of a special region of an image. In all these papers the image is categorized to the region of important (ROI) and the region of background (ROB). In paper~\cite{lin2011roi}, the ROI and ROB are determined manually. But, the papers~\cite{li2016semi, li2016semi} are used for protection of face or faces in the image and the detection of faces are automatically using face detection algorithms. Recovery watermark is calculated from the ROB region and it is embedded into the ROI region. Quality and robustness of these methods are high because the protection is just done on the ROI region.

As seen in the related works of this section, the amount of extracted bits for all regions of the image is equal. Although, some papers such as~\cite{lin2011roi, li2016semi, cruz2015face} have paid attention to the amount of extracted bits for the special ROI regions. Aside from the amount of extracted bits for different regions, the embedding regions is another important problem that is not taken into consideration by the state-of-the-art methods. So, in this paper, the original image is categorized into three regions automatically based on the texture of the image. The amount of extracted bits for each block in different regions is various and the extracted bits for each block in a region is embedded into a different block in its region. Also, a two-sided circular block dependency is proposed in front of the one-sided circular block dependency in~\cite{li2015semi}, which the two-sided circular block dependency has some advantages over the one side circular block dependency.

\section{Proposed Method} \label{lbl.proposedmethod}
\begin{figure*}[!htbp]
  \centering
  \includegraphics[width=0.9\textwidth]{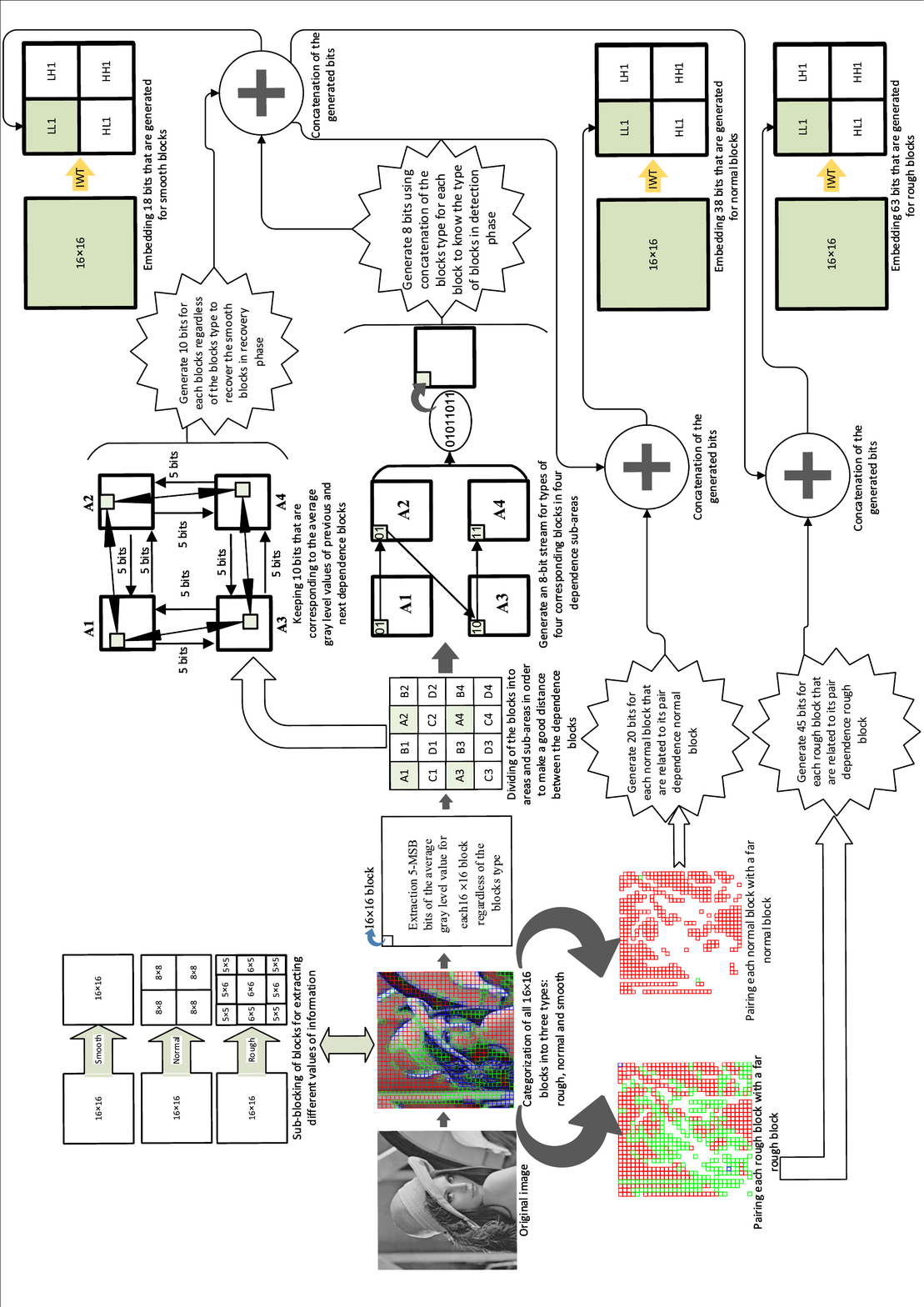}
  \caption{Diagram of the information extraction and the embedding phase.}\label{fig.DiagramEN}
\end{figure*}

The proposed algorithm is made up of four subsections. Information extraction and embedding algorithms are described in  subsection~\ref{subsec.InformationExtractionAndEmbedding}. Then, the tamper detection and tamper recovery algorithms are described in subsections~\ref{subsec.TamperDetetcion} and~\ref{subsec.TamperDetetcion}, respectively. Diagram of the information extraction and the embedding phase of the proposed method are shown in Fig.~\ref{fig.DiagramEN}, which each part of it will be explained in detail in the sub-sections.

\subsection{Information Extraction and Embedding}\label{subsec.InformationExtractionAndEmbedding}
\begin{figure}[h]

        \centering

 \begin{subfigure}[b]{0.47\columnwidth}
                \centering
                \includegraphics[width=\textwidth]{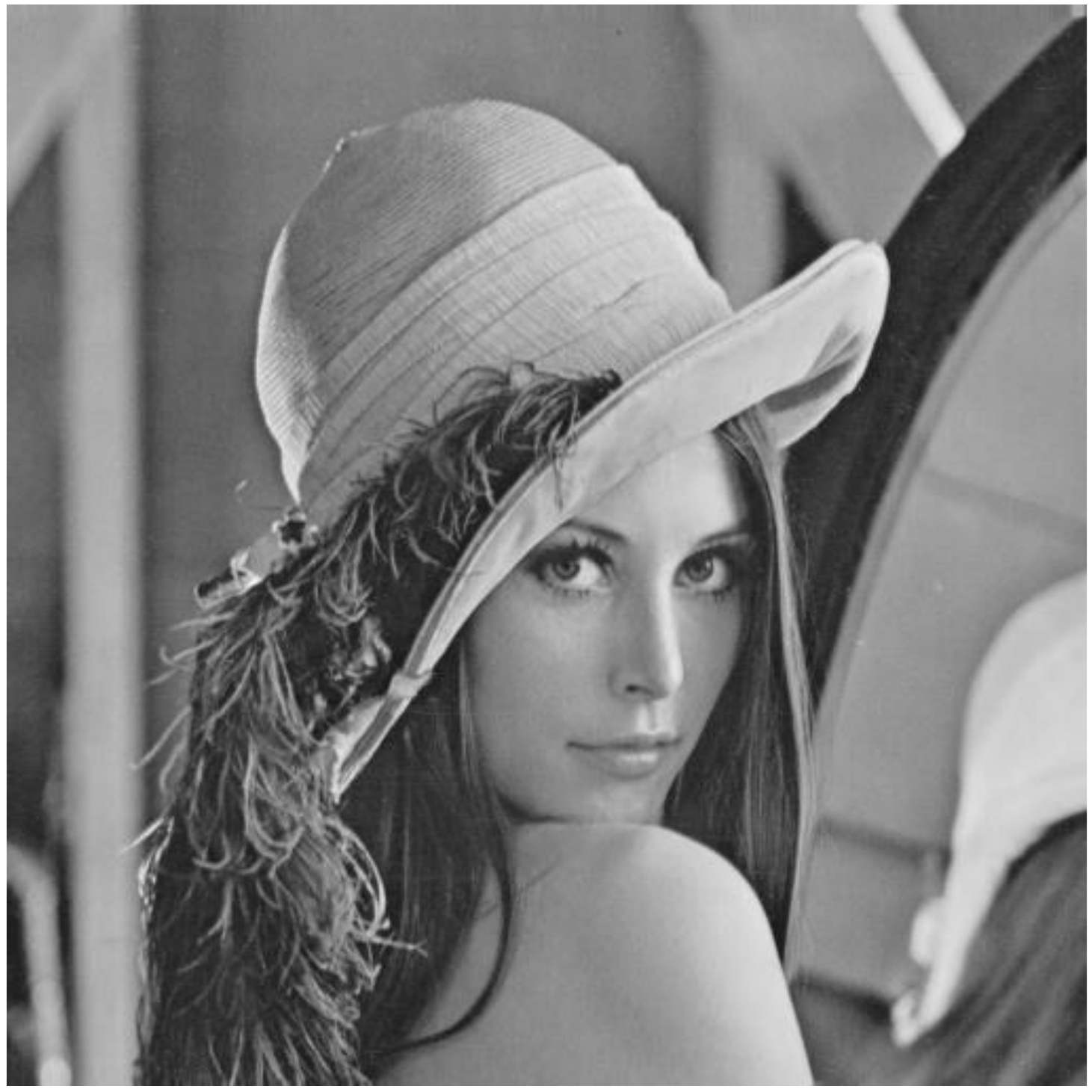}
                \caption{Original Image}
                \label{subfig.aOriginal}
        \end{subfigure}
        \begin{subfigure}[b]{0.49\columnwidth}
                \centering
                \includegraphics[width=\textwidth , trim=0 10 0 0,clip=true]{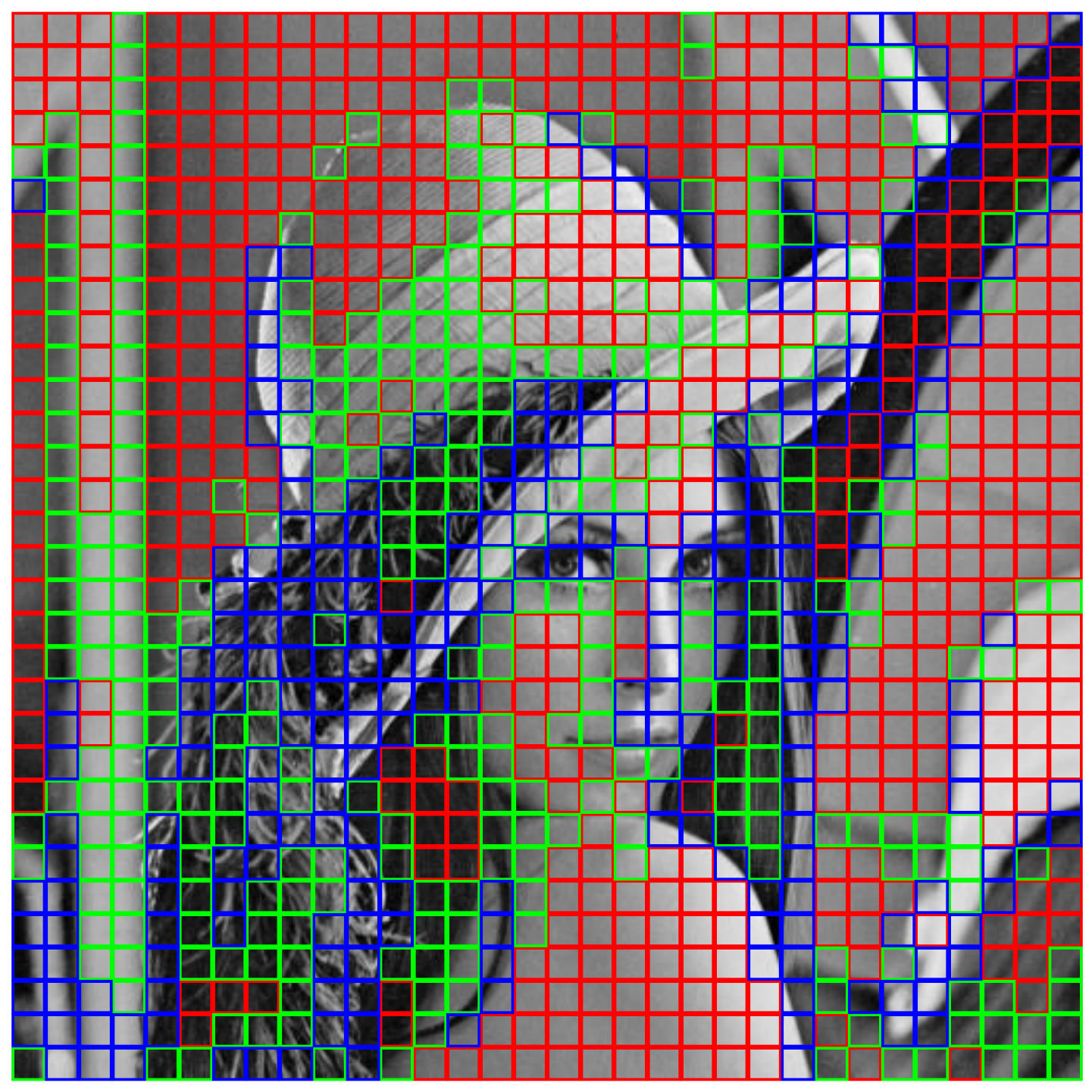}
                \caption{Categorazation of blocks}
                \label{subfig.bCategoraized}
        \end{subfigure}
                \caption{Categorization of image blocks using standard deviation. Fig.\ref{subfig.aOriginal} is the original image and Fig.\ref{subfig.bCategoraized} is the categorized blocks. The Red blocks are smooth, the Green blocks are normal and the Blue blocks are rough. }
                \label{fig.Categorization}
\end{figure}

In this method, the original image is divided into 16$\times$16 non-overlapping blocks and the standard deviation of each block is calculated. Then, all the standard deviations are normalized and they are categorized into three types using two experimental thresholds~$Th_1$ and~$Th_2$. For normalization of the standard deviation values the Eq.\ref{eq.Normalization} is used, which $X$, $X_{min}$ and $X_{max}$ are the current standard deviation, minimum value and maximum value between all of the standard deviation values, respectively. In this situation, all blocks of the image are categorized into one of the smooth, normal or rough types. Categorization result of blocks according to the texture is shown in Fig.\ref{fig.Categorization}.

\begin{equation}\label{eq.Normalization}
{\footnotesize
Norm(X)=\frac{X-X_{min}}{X_{max}-X_{min}}.
  }
\end{equation}

\begin{figure}[h]
  \centering
  \includegraphics[width=0.6\columnwidth]{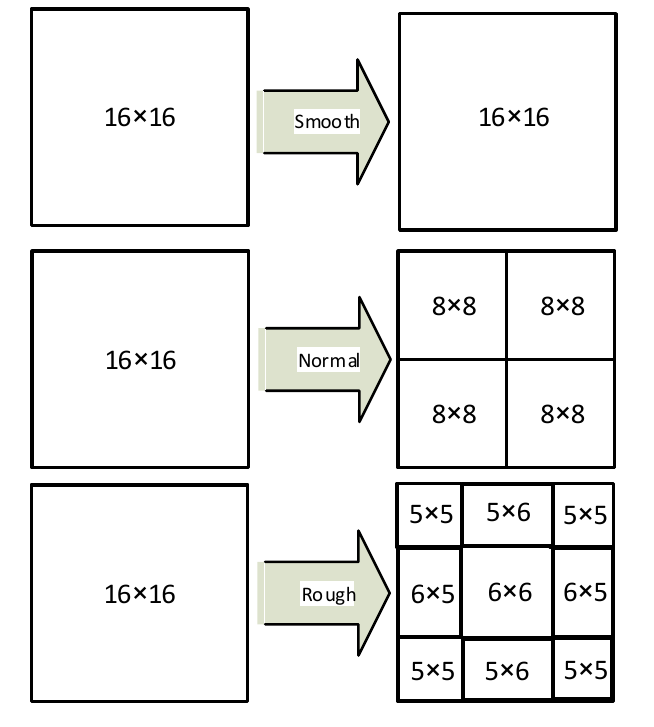}
  \caption{The smooth, normal and rough blocks are divided into 1, 4 and 9 sub-blocks, respectively.}\label{fig.subblocking}
\end{figure}

In this method, each smooth, normal and rough block is divided into 1, 4 and 9 sub-blocks, respectively. Then, the average value of gray level for each sub-block is calculated. Also, regardless of the type of blocks, the average value of gray level is calculated for all $16\times16$ blocks and five the most significant bits (5-MSB) of the average values are calculated for use in embedding phase. The block dividing operation into its sub-bands is shown in Fig.\ref{fig.subblocking}.

Pairwise block dependency is used for the blocks whose type is normal or rough. It is important that the information of a normal block must be embedded into another normal block and the information of a rough block must be embedded into another rough block. As seen in Fig.\ref{fig.subblocking}, each normal and rough block must maintain 20 bits and 45 bits, respectively, for its pair dependent block. The pair dependent blocks must be far from each other. For this purpose, a random block is selected from normal blocks and then the furthest block to the selected block is found, which this far block is not paired with any other blocks. The similar operation must be done for all normal and rough blocks.

\begin{figure}[h]
  \centering
  \includegraphics[width=0.3\textwidth]{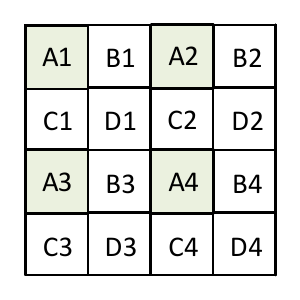}
  \caption{Block distance structure for providing a sufficient distance between the dependent blocks. For a $512\times512$ image, the $A1$ sub-area contains 64 ($8\times8$) blocks, which each block contains 256 ($16\times16$)  pixels.}\label{fig.SufficientDistance}
\end{figure}

As mentioned before, 5-MSB bits is calculated for all $16\times16$ blocks and these bits will be used for tamper detection in all blocks and tamper recovery in the smooth blocks. According to this fact that these 5-MSB bits of the average values are important in tamper localization, so the blocks must have a sufficient distance with their dependent blocks. For this purpose, a block distance structure is proposed, which is shown in Fig.\ref{fig.SufficientDistance}. Using this block distance structure, each block has a minimum and maximum distance with its dependent blocks. In this structure, four main areas ($A$, $B$, $C$ and $D$) are considered for the original image, which each area have four dependent sub-areas.

\begin{figure}[h]
  \centering
  \includegraphics[width=0.4\textwidth]{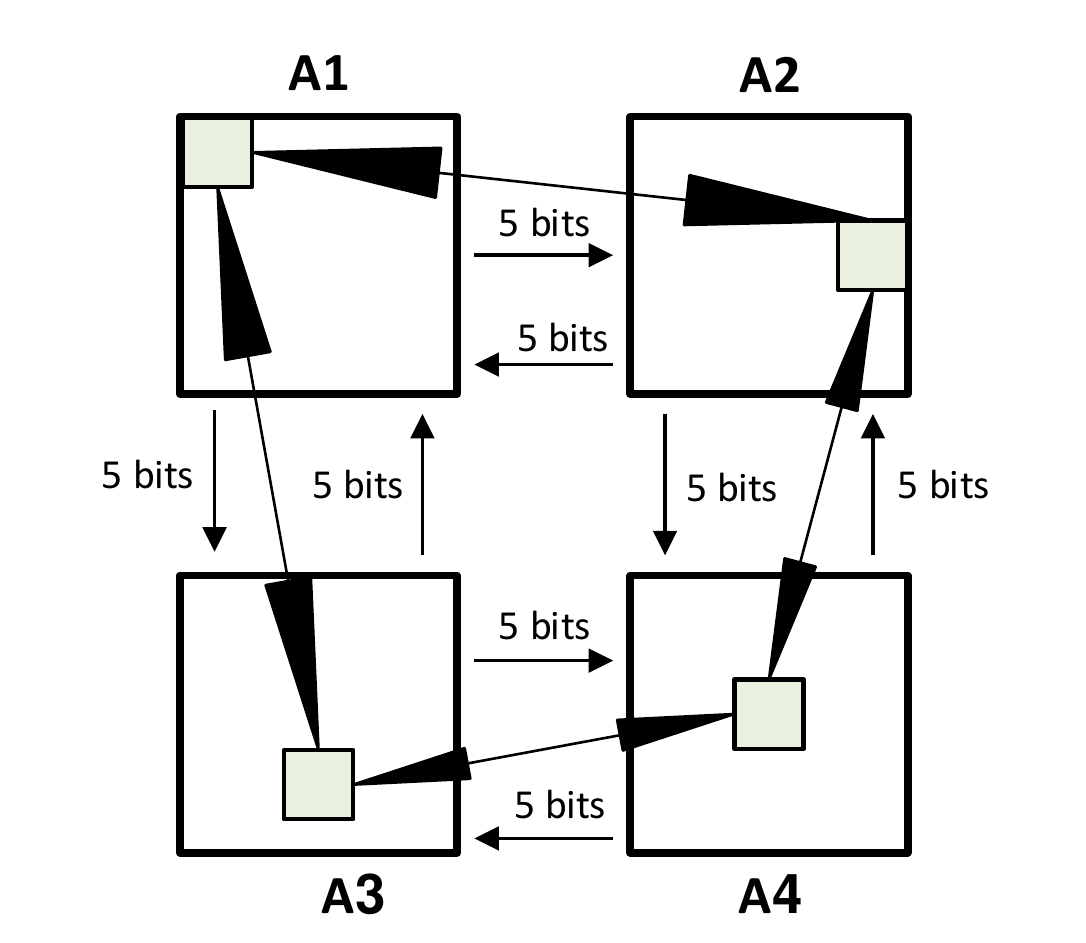}
  \caption{ Two sided circular block dependency between sub-area blocks of area $A1$.}\label{fig.TwoSideCircularBlockDependency}
\end{figure}

 For tamper localization, each random block of  $A1$ sub-area keeps 5-MSB bits from a random block of $A4$ sub-area and 5-MSB bits from a random block of  $A2$ sub-area. The $A4$ and  $A2$ are previous and next dependent sub-areas with  $A1$ sub-area. The two-sided circular block dependency between blocks is shown in Fig.\ref{fig.TwoSideCircularBlockDependency}. There is similar block dependency for other blocks in other sub-areas in B, C, and D  areas. This dependency is more powerful than the pairwise block dependency and the one side circular block dependency, in the tamper localization phase.

\begin{figure}[h]
  \centering
  \includegraphics[width=0.5\textwidth]{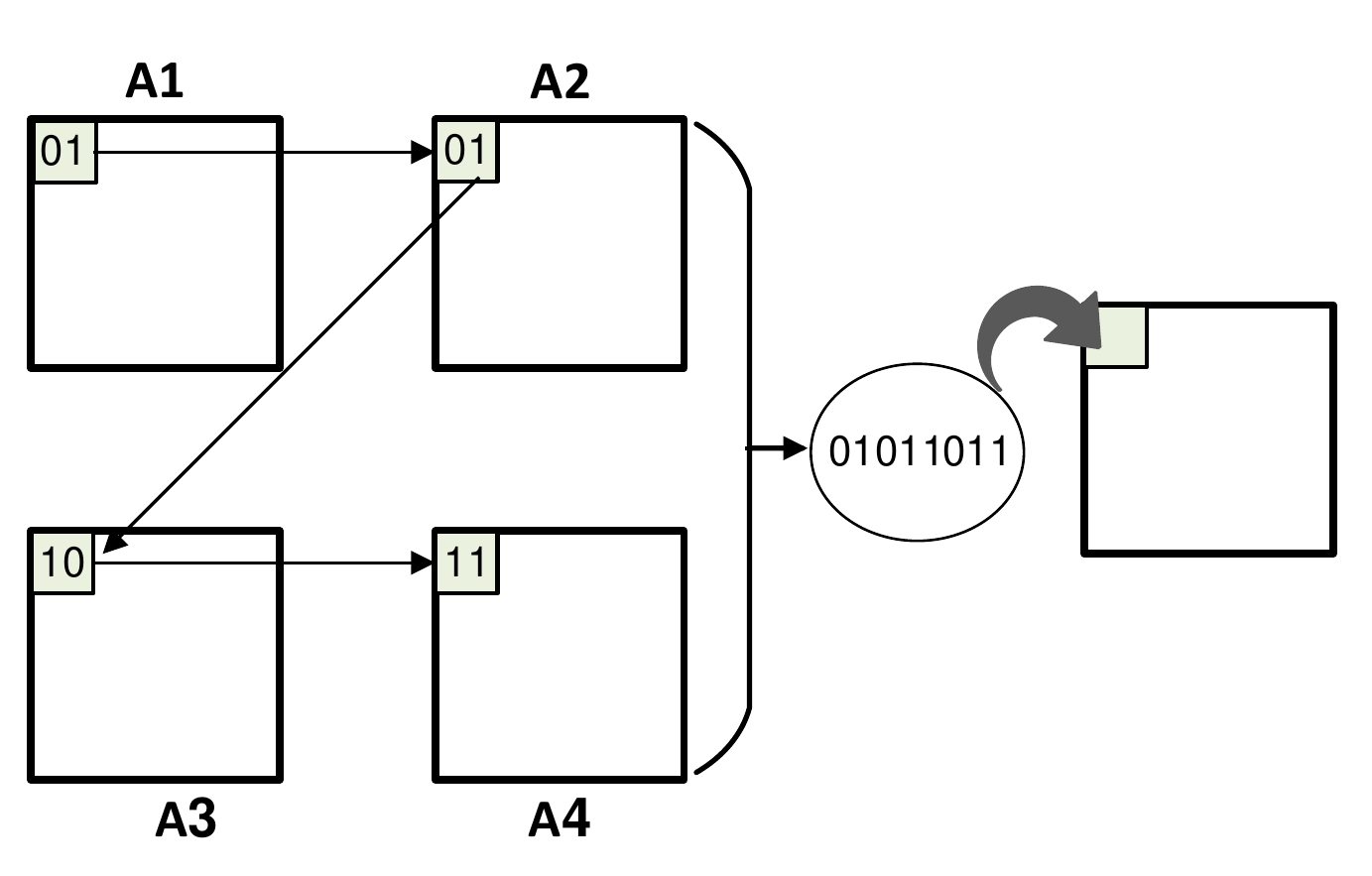}
  \caption{ Concatenation of block types ("01" for a smooth block, "10" for a normal block and "11" for a rough block) for corresponding blocks of four sub-areas. }\label{fig.TypeCreation}
\end{figure}

This method needs to know the type of each block in the tamper detection and recovery phases. So, the smooth, normal and rough block types are shown by binary bits "01", "10" and "11", respectively. In this method, the block type of four corresponding blocks, in four related sub-areas, are concatenated and 8 bits are created for them. Then, four copies of these 8 bits are embedded in the random location of the four related sub-areas. These four copy of block types will be used in a voting mechanism in tamper detection and recovery phases. The 8 bits creation operation is shown in Fig.~\ref{fig.TypeCreation}.

So far, each block (with smooth, normal or rough type) must maintain 10 bits for two side circular block dependency and 8 bits for determining the block's type. Furthermore, normal and rough blocks need to maintain 20 and 45 bits, respectively for their pairwise dependency. Thus, each smooth, normal and rough block needs to maintain 18, 38 and 63 bits, respectively.

The calculated bits for each block is embedded in first approximation sub-band (LL1) of Integer Wavelet Transform (IWT) of each block using Quantization Index Modulation (QIM) method that are shown in relations~\ref{eq.CbarN}~-~\ref{eq.v2}. For embedding in each block, 18, 38 and 63 coefficients of LL1 sub-band are selected randomly for smooth, normal and rough blocks, respectively.  In these relations $C_{n}$ is the selected coefficient for embedding and  $\tilde{C_{n}}$ is the embedded coefficient. Also, $w_n$ is one of the watermark bits. After any tamper in the watermarked image, the proposed algorithm needs to find the tampered blocks and the type of all blocks and finally recovering the tampered blocks, which these phases are described in the following subsections.

\begin{equation}\label{eq.CbarN}
\mathsmaller{
{\footnotesize
\tilde{C_{n}}
=\begin{cases}
  v_1 , & \mbox{if } |C_{n}-v_1|\leq |C_{n}-v_2| ,\\
  v_2 , & \mbox{otherwise}.
  \end{cases}
  }
  }
\end{equation}

\begin{equation}\label{eq.v1}
\mathsmaller{
{\footnotesize
v_1
=\begin{cases}
 2S\lfloor \frac{C_{n}}{2S}  \rfloor , & \mbox{if } w_n==0 ,\\
 2S\lfloor \frac{C_{n}}{2S}  \rfloor + S , & \mbox{if }w_n==1.
  \end{cases}
  }
  }
\end{equation}

\begin{equation}\label{eq.v2}
\mathsmaller{
{\footnotesize
v_2
=v_1+2S.
  }
  }
\end{equation}

\subsection{Tamper Detection}\label{subsec.TamperDetetcion}
As mentioned in the previous section, 10 bits are embedded in each block for previous and next dependent blocks using two side circular block dependency. These 10 bits consists of 5-MSB bits for the average gray level value of the previous dependent block and 5-MSB bits for the average gray level value of next dependent block. Relation~\ref{eq.extraction} is used in order to extract these 10 bits for each block from the coefficients of LL1 sub-band.

\begin{equation}\label{eq.extraction}
\mathsmaller{
{\footnotesize
\tilde{w_n}
=\begin{cases}
 0  , & \mbox{if } \textbf{round} (\frac{\tilde{C_{n}}}{S})==\textbf{even} ,\\
 1 , & \mbox{if } \textbf{round}(\frac{\tilde{C_{n}}}{S})==\textbf{odd}.
  \end{cases}
  }
  }
\end{equation}

In this step, the status of each block is defined by healthful block, fully destroyed block or partially destroyed block. The status of a block is healthful if the 5-bits generated from the average gray level value of current block be extractable from at least one of the previous or next dependent blocks and also the 5-MSB bits generated from the average gray level value of the previous and next dependent blocks be extractable from the current block. The status of a block is fully destroyed if the 5-bits generated from the average gray level value of current block not be extractable from both of the previous and next dependent blocks and also the 5-MSB bits generated from the average gray level value of the previous and next dependent blocks not be extractable from the current block.

\begin{figure}[h]
  \centering
  \includegraphics[width=0.45\textwidth]{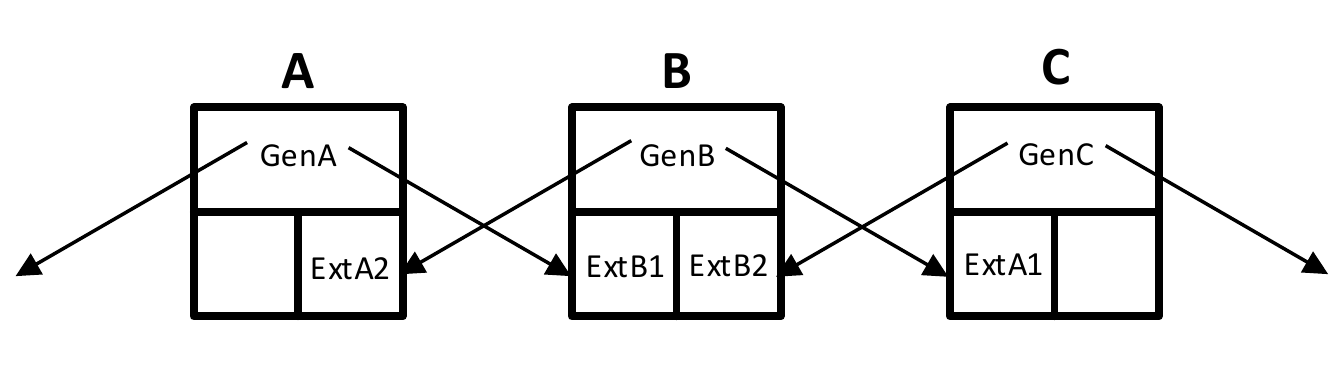}
  \caption{Making decision operations on the status of exemplar block $B$. }\label{fig.ExtGen}
\end{figure}

The status of a block is partially destroyed if the 5-MSB bits generated from the average gray level value of the previous and next dependent blocks not be extractable from the current block and also the 5-bits generated from the average gray level value of current block be extractable from at least one of the previous or next dependent blocks. The pairwise block dependency or the one-sided circular bock dependency in~\cite{li2015semi} are not able to distinguish the partially destroyed blocks. This type of distinguished blocks increases the accuracy of tamper localization and recovery algorithms. Making decision operations on the status of exemplary block $B$ in Fig.\ref{fig.ExtGen} , are shown in relations~\ref{lbl.Healthful}~-~\ref{lbl.PartiallyDestroyed}. In these relations, the meaning of $Gen_B$ is 5-MSB bits generated from the average gray level value of current block $B$. Also, $Ext_{B1}$ and $Ext_{B2}$ are the extracted 5-MSB information bits from block $B$ that are corresponding to the average gray level value of block $A$ and $C$ in embedding phase, respectively.

\begin{equation}\label{lbl.Healthful}
\begin{array}{c@{\qquad}c}
\begin{aligned}
\textrm{\textbf{Status(B) is Healthful if: } } {} & \\ \{ \left(Gen_A==Ext_{B1}\right) \| \left( Gen_C==Ext_{B2} \right)\}
\end{aligned}
\end{array}
\end{equation}

\begin{equation}\label{lbl.FullyDestroyed}
\begin{aligned}
\textbf{\textrm{Status(B) is Fully Destroyed if:}  } {} & \\ \{ \left( Gen_A\neq Ext_{B1}\right) \&\left( Gen_C\neq Ext_{B2}\right)\\
      & \\ \&\left( Gen_B\neq Ext_{A2}\right) \&\left( Gen_B\neq Ext_{C1}\right)\}
\end{aligned}
\end{equation}

\begin{equation}
\begin{aligned}\label{lbl.PartiallyDestroyed}
\textrm{\textbf{Status(B) is Partially Destroyed if: } } {} & \\ \{ \left[\left( Gen_A\neq Ext_{B1}\right) \&\left( Gen_C\neq Ext_{B2}\right)\right]\\
      & \\ \& \left[ \left( Gen_B==Ext_{A2}\right) \| \left( Gen_B==Ext_{C1}\right)\right]\}
\end{aligned}
\end{equation}

 \begin{algorithm}[h]
     \SetAlgoLined
     \KwIn{Original image}
     \KwOut{Watermarked image}
    Dividing image into non-overlapping $16\times 16$ blocks\;
     Calculation of standard deviation for each block and normalize all of them\;
     Categorization of blocks into smooth, normal or rough block types using two thresholds $Th_1$ and $Th_2$ on the normalized standard deviation values (Fig.~\ref{fig.Categorization})\;
     Dividing the smooth, normal and rough blocks into 1, 4, and 9 sub-blocks, respectively (Fig.~\ref{fig.subblocking})\;
     Extracting 5-MSB bits from the average gray value of each sub-block and making 20 and 45 bits for normal and rough blocks, respectively\;
      Extracting 5-MSB bits of the average gray level value for all $16\times16$ blocks\;
      Concatenation the 5-MSB bits of  the previous and next dependent blocks for each block (Fig.~\ref{fig.TwoSideCircularBlockDependency})\;
      Generating 8 bits from four corresponding blocks type for each block (Fig.~\ref{fig.TypeCreation})\;
      Finding a far pair block for each normal and rough block with the same type\;
      making 18, 38 and 63 bits using the results of steps 5, 6 and 7 for each smooth, normal and rough block, respectively.\;
      Embedding the extracted bits for each block in the $LL_1$ sub-band of IWT transform of block using the QIM method in Eq.~\ref{eq.CbarN}~-~\ref{eq.v1}.

     \caption{Information extraction and embedding algorithm }
     \label{algo.InfExtractionEmbedding}
     \end{algorithm}

\subsection{Make Decision On Blocks Type}\label{subsec.MakeDecisionOnBlockTypes}

As mentioned in subsection~\ref{subsec.InformationExtractionAndEmbedding}, for four corresponding blocks in four sub-areas $A1$, $A2$, $A3$ and $A4$, eight bits of types are concatenated and four copy of these eight bits are embedded in random locations of four sub-areas $A1$, $A2$, $A3$ and $A4$. Type detection of each block is done by voting on the four copy of eight bits. There are two states for voting. The first state is for partially or fully destroyed blocks and the second state is for healthful blocks. In the first state, the votes of other three blocks are gathered for current block type decision. In the second state, the vote of the current block and two other healthful blocks in other three sub-areas are used for current block type decision. So, these two states try to use just healthy blocks in making the decision on the block's type.

\subsection{Tamper Recovery}\label{subsec.TamperRecovery}
After detecting the type of the exact block and tampered blocks, there is need to extract the embedded information bits of smooth, normal and rough blocks. As seen from the previous subsections the information bits of smooth blocks are embedded in the previous and next dependent blocks using the two side circular block dependency. For a tampered smooth block, its corresponding information bits must be calculated from a not destroyed dependent block. Also, the paired block of each normal or rough block is in a random far location with the same block type. So, in this step, the paired block of each normal and rough block must be found and then the embedded information bits (20 and 45 bits for normal and rough blocks, respectively) must be extracted. As mentioned in the subsection~\ref{subsec.InformationExtractionAndEmbedding}, the information bits for each block are embedded in the LL1 sub-band of the corresponding block IWT transform and must be extracted using the relation~\ref{eq.extraction}.

So, in this step, 20 bits are calculated for four sub-blocks of each normal block and  45 bits are calculated for nine sub-blocks of each rough block. Every five bits of 20 or 45 bits are the 5-MSB bits of the average gray level value of a sub-block. So, the binary value "100" is appended to the LSB bits of these 5-MSB bits in order to form an eight bits binary value and the decimal value of the eight bits binary value is used as the average gray level value of each sub-block.

\begin{figure}[h]
  \centering
  \includegraphics[width=0.35\textwidth]{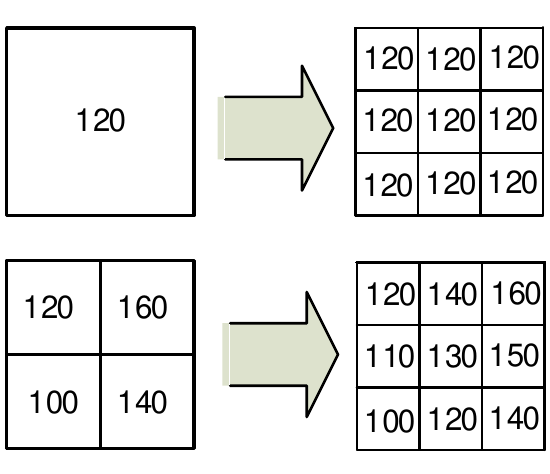}
  \caption{ Reforming of the smooth and normal blocks to the rough block form. }\label{fig.ExpandBlk}
\end{figure}

Each smooth, normal and rough block is consist of 1, 4 and 9 decimal values, respectively. In this method, for recovery of tampered blocks, the cubic interpolation is applied to the calculated decimal values for sub-blocks. In order to increase the benefits of cubic interpolation, the contents of smooth and normal blocks are reformed to the form of the rough block as shown in Fig.\ref{fig.ExpandBlk}. The reformation and cubic interpolation are done on the whole of tampered image blocks. These operations combine the neighbor sub-blocks and blocks, which decrease the blocking effect in tampered regions.


 \begin{algorithm}[h]
     \SetAlgoLined
     \KwIn{Tampered image}
     \KwOut{Recovered image}
     Dividing image into non-overlapping $16\times 16$ blocks\;
     Extracting 8 bits that are related to the type of four correspond blocks from $LL_1$ sub-band of IWT transform using Eq.~\ref{eq.extraction}\;

     Applying a voting algorithm on the block types for make decision on the type of image blocks (subsection~\ref{subsec.MakeDecisionOnBlockTypes})\;

Extracting 10 bits that are related to the previous and next dependent block using Eq.~\ref{eq.extraction}\;

Making decision on the status of each block using Eq.~\ref{lbl.Healthful}~-~\ref{lbl.PartiallyDestroyed}\;
Finding the coressponding far pair block for each normal and rough block\;

Extracting 20 and 45 bits from  the corresponding far pair block for each normal and rough block using Eq.~\ref{eq.extraction}\;

Converting the form of smooth and normal blocks to the form of a rough block (Fig.~\ref{fig.ExpandBlk})\;

Applying cubic interpolation on the values of image sub-blocks to become equal to the size of original image\;

replacing the tampered blocks with the results of step 9.
     \caption{Tamper detection and recovery algorithm }
     \label{algo.TamperDetectionRecovery}
     \end{algorithm}

\section{Experimental Results}\label{lbl.experimentalresults}

\begin{figure*}[!htbp]
  \centering
  \includegraphics[width=0.7\textwidth]{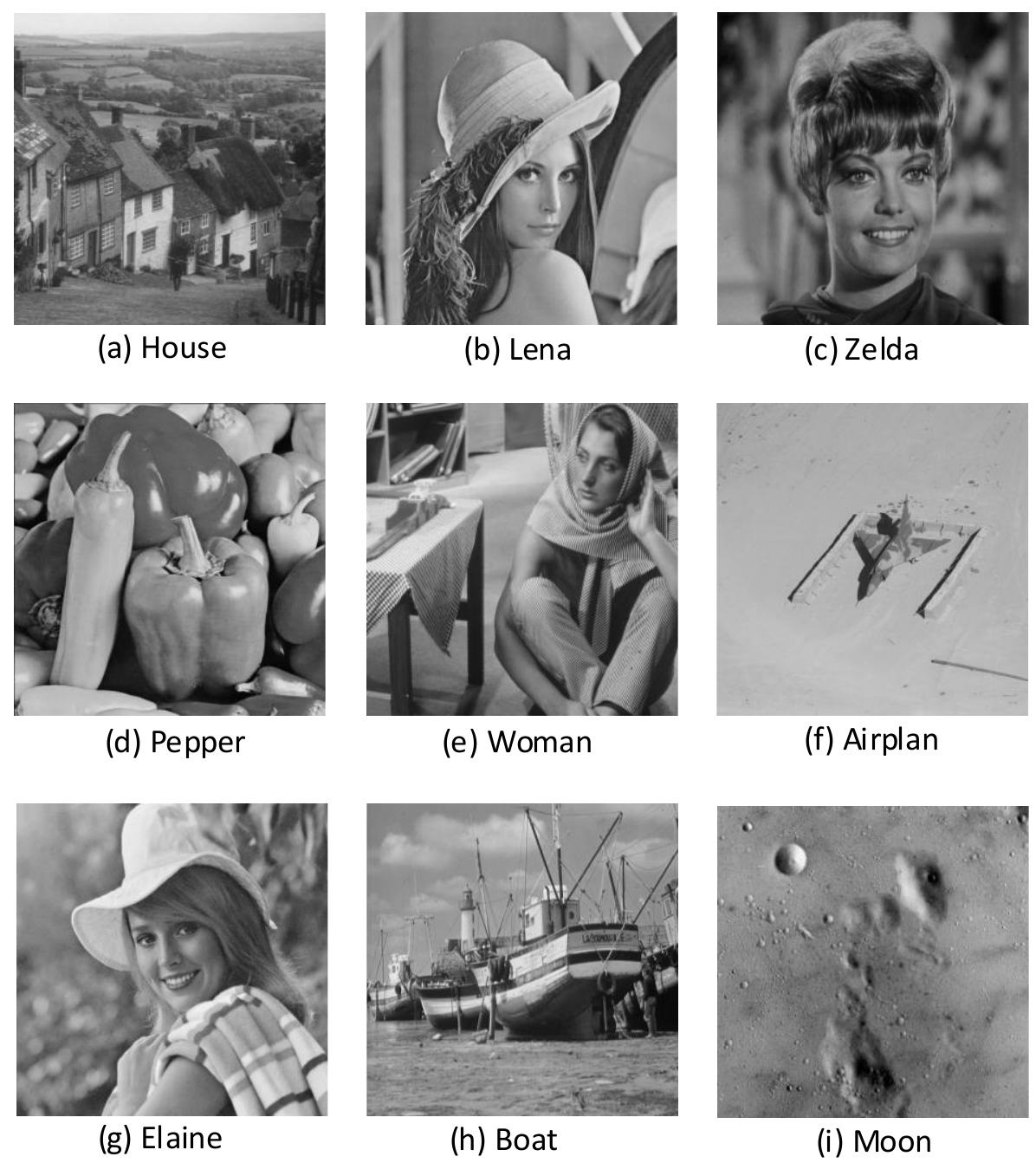}
  \caption{ Some of the standard images of USC-SIPI image database~\cite{USC}. }\label{fig.StandardImages}
\end{figure*}

For evaluation of the proposed method twenty $512\times 512$  standard gray images are used, which are selected from the USC-SIPI image database~\cite{USC}. Some of the standard images of this database are shown in Fig.~\ref{fig.StandardImages}. For evaluation of robustness, the JPEG and JPEG2000 compressions and copy-move attacks are applied to the watermarked image.

Peak Signal to Noise Ratio (PSNR) is used for evaluation of the visual quality of the watermarked image and the recovered image. This criterion compares pixel by pixel similarity between the original image, watermarked image and recovered image and is defined as Eq.~\ref{eq.psnr}.
\begin{equation}\label{eq.psnr}
  \textrm{PSNR}(f,f_w)=
  10\log_{10}\left [{\frac{\max_{\forall(m,n)}f^2(m,n)}{\frac{1}{N_f}\sum_{\forall(m,n)}{\left( f_w(m,n)-f(m,n) \right)}^2}}\right].
\end{equation}

In Eq.~\ref{eq.psnr}, $f(m,n)$ is the original image, $f_w(m,n)$ is the watermarked (or recovered) image and $N_f$ is the number of pixels in image.

Another visual quality measure that is used in this paper is Structural Similarity (SSIM), which the structure of the image is considered in it. The values of SSIM measure is in the range [0,1], which 0 is the minimum similarity and 1 is the maximum similarity.

 \begin{figure*}[!htbp]
  \centering
  \includegraphics[width=0.9\columnwidth]{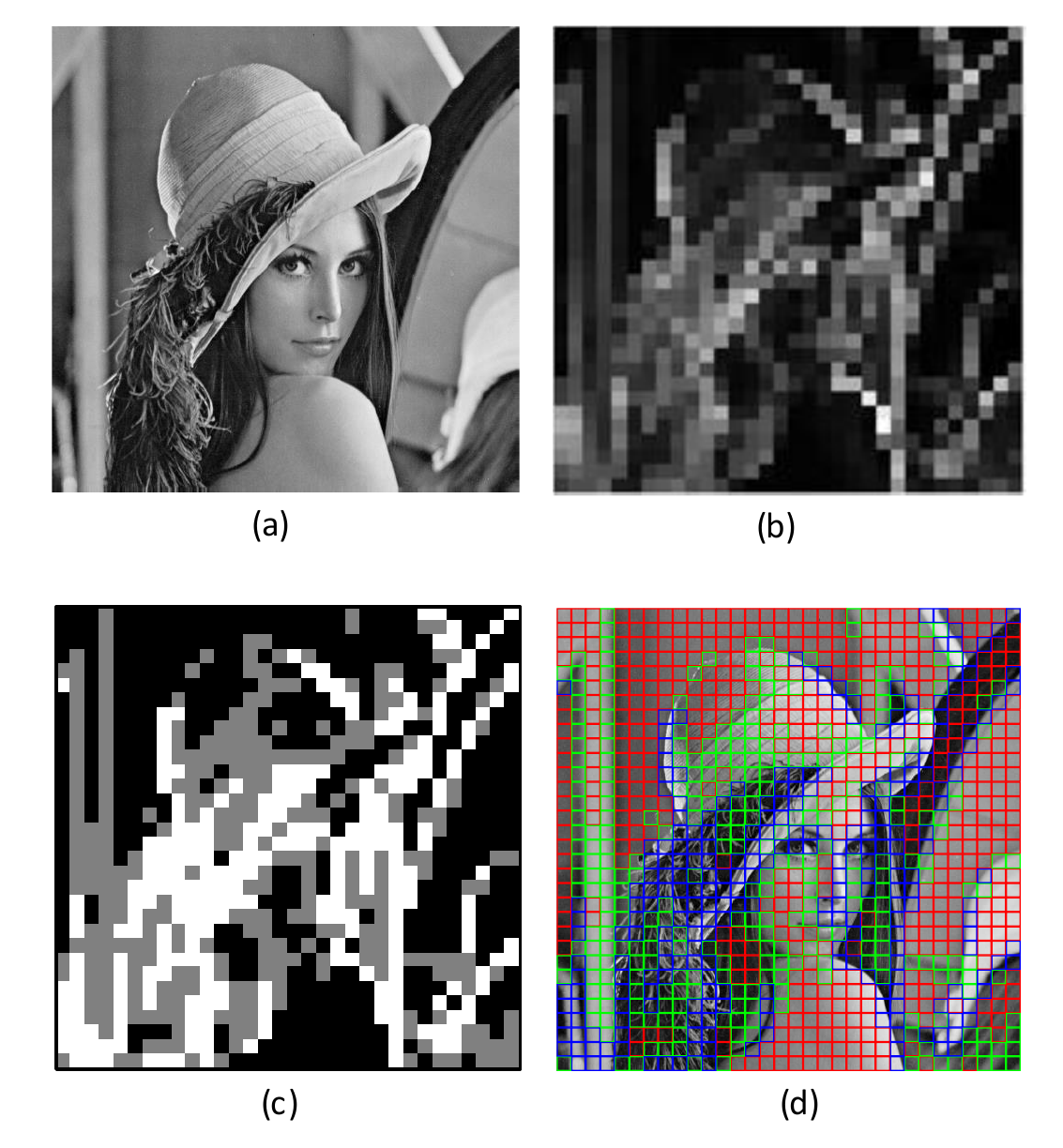}
  \caption{ categorization of blocks. (a) Original image. (b) Standard deviation values. (c) Categorization using thresholds $Th_1=0.1$ and $Th_2=0.3$ (black color for smooth, gray color for normal and white color for rough block). (d) Representation of each block on image (red color for smooth, green color for normal and blue color for rough block). }\label{fig.ResultsCategorization}
\end{figure*}

\begin{figure}[H]%
\centering
\begin{subfigure}{0.24\columnwidth}
\includegraphics[width=\textwidth]{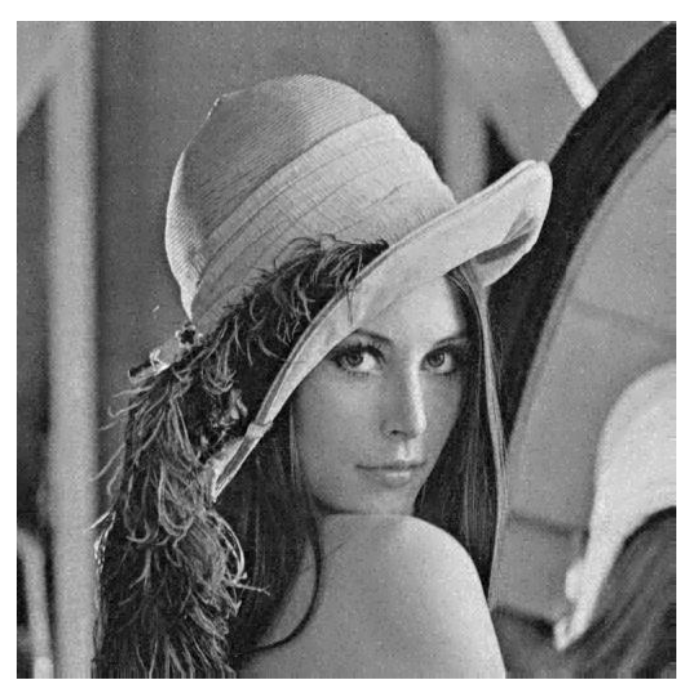}%
\caption{}%
\label{subfig.lena.a}%
\end{subfigure}\hfill%
\begin{subfigure}{.24\columnwidth}
\includegraphics[width=\textwidth]{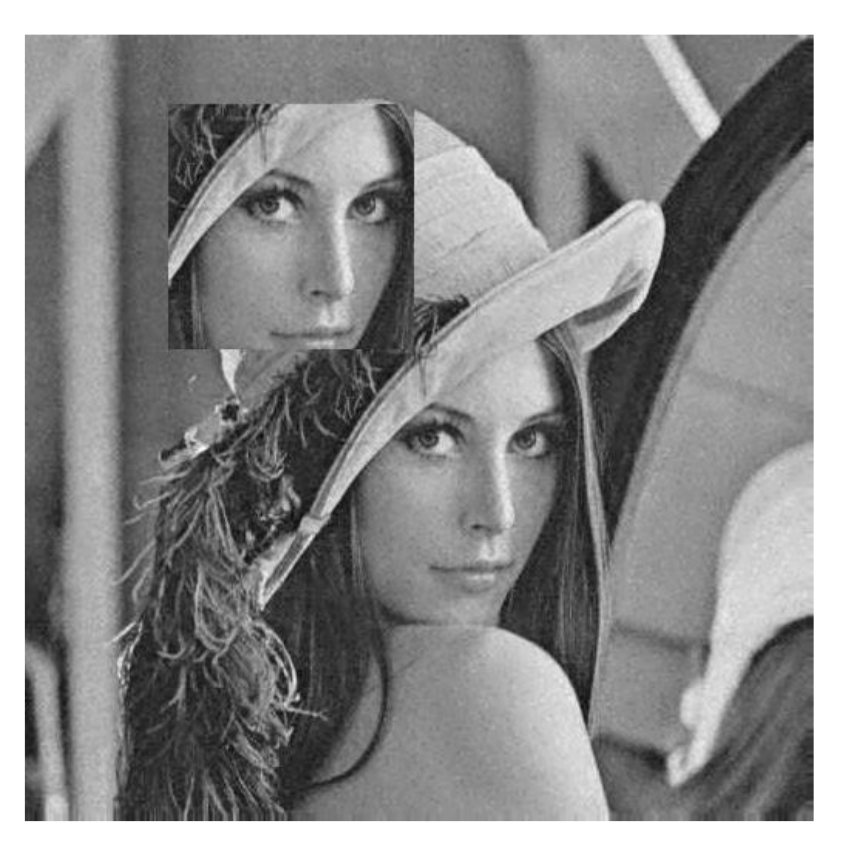}%
\caption{}%
\label{subfig.lena.b}%
\end{subfigure}\hfill%
\begin{subfigure}{.24\columnwidth}
\includegraphics[width=\textwidth]{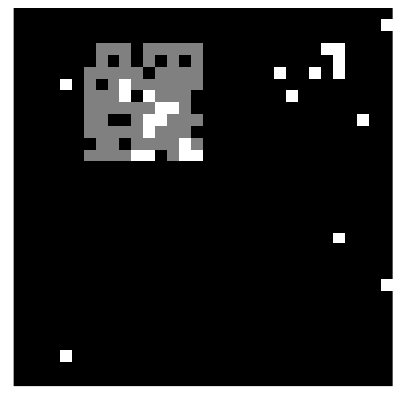}%
\caption{}%
\label{subfig.lena.c}%
\end{subfigure}\hfill%
\begin{subfigure}{.24\columnwidth}
\includegraphics[width=0.98\textwidth]{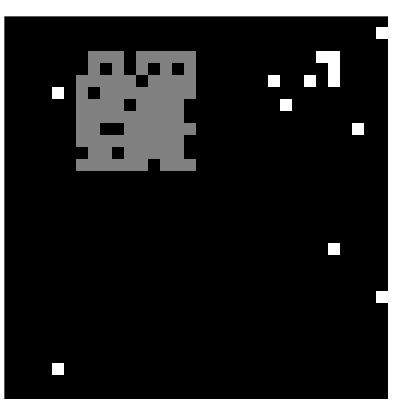}%
\caption{}%
\label{subfig.lena.d}%
\end{subfigure}\hfill%
\begin{subfigure}{.24\columnwidth}
\includegraphics[width=\textwidth]{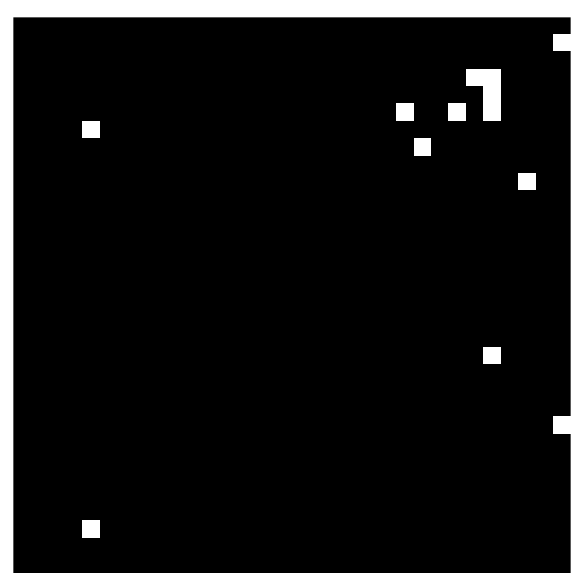}%
\caption{}%
\label{subfig.lena.e}%
\end{subfigure}\hfill%
\begin{subfigure}{.24\columnwidth}
\includegraphics[width=\textwidth]{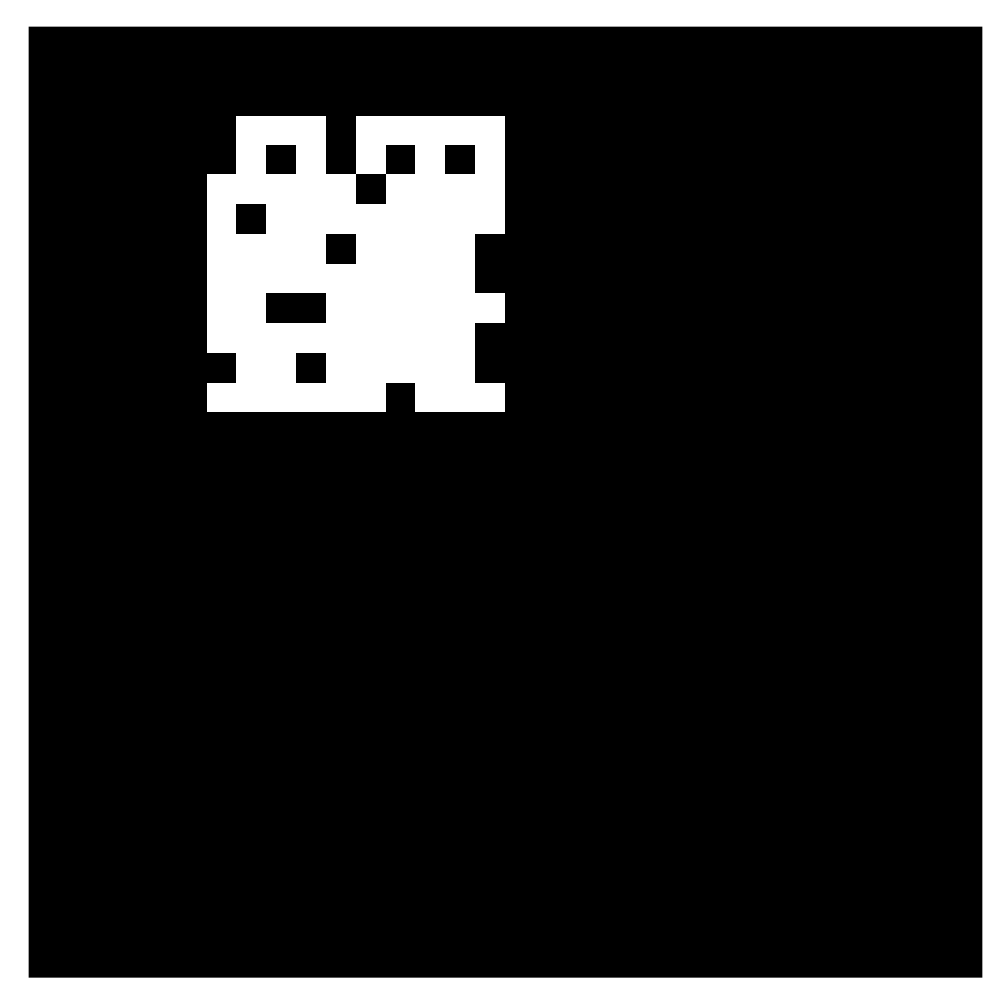}%
\caption{}%
\label{subfig.lena.f}%
\end{subfigure}\hfill%
\begin{subfigure}{.24\columnwidth}
\includegraphics[width=\textwidth]{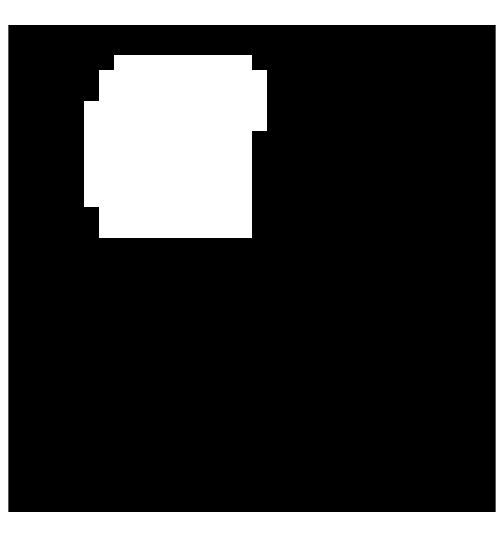}%
\caption{}%
\label{subfig.lena.g}%
\end{subfigure}%
\label{Fig.LentaTamperDetetction}
\caption{Tamper detection steps on the Lena standard image. ّFig.\ref{subfig.lena.a} is the watermarked image with PSNR=35.20 dB. Fig.\ref{subfig.lena.b}, shows the tampeded and JPEG compressed image with QF=50. Fig.\ref{subfig.lena.c}, shows the three detected block types (black is the smooth block, gray is the fully destroyed block, white is the rough block). Fig.\ref{subfig.lena.d}, shows the conversion of partially destroyed block to fully destroyed block if they have continuty. Fig.\ref{subfig.lena.e}, shows the blocks that do not have any role in the continuing tamper detection process. Fig.\ref{subfig.lena.f}, shows the continuous partially destroyed blocks and fully destroyed blocks independently. Fig.\ref{subfig.lena.f}, shows the tamper deteted blocks after filling the hole and contour blocks (FR=0\% and FA=3.2\%). } \label{fig.TamperDetection1}
\end{figure}

\begin{figure}[h]%
\centering
\begin{subfigure}{0.24\columnwidth}
\includegraphics[width=\textwidth]{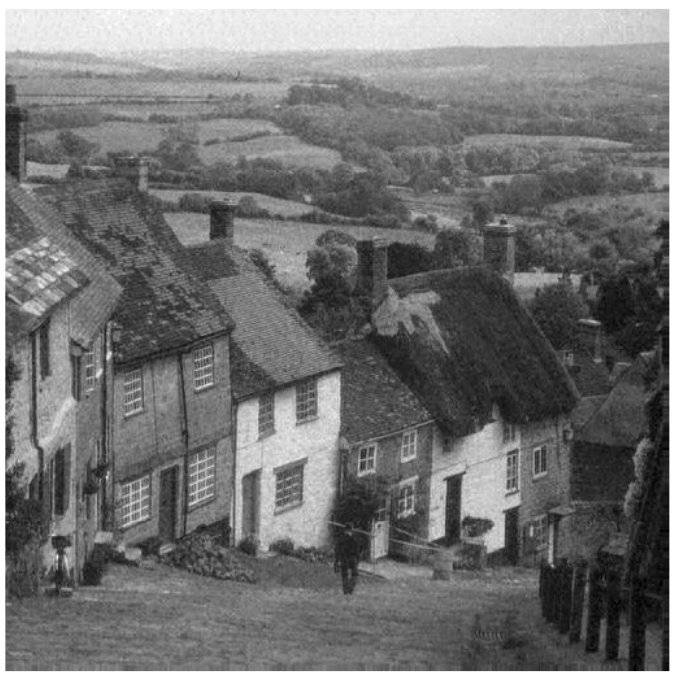}%
\caption{}%
\label{subfig.house.a}%
\end{subfigure}\hfill%
\begin{subfigure}{.24\columnwidth}
\includegraphics[width=\textwidth]{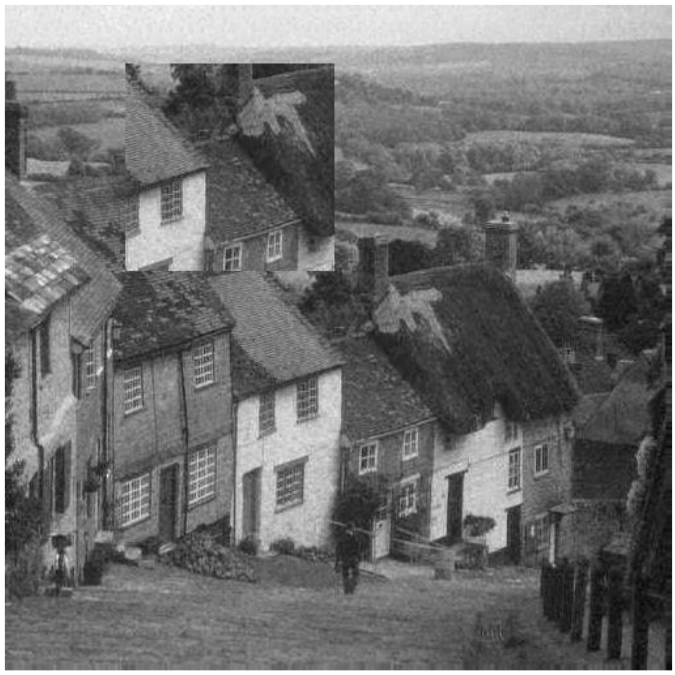}%
\caption{}%
\label{subfig.house.b}%
\end{subfigure}\hfill%
\begin{subfigure}{.24\columnwidth}
\includegraphics[width=\textwidth]{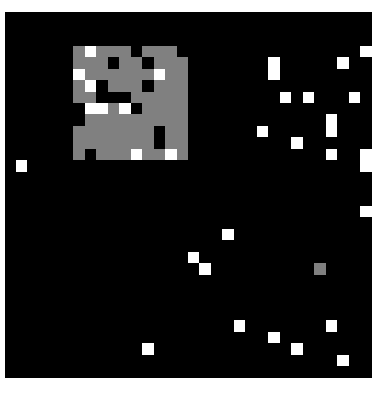}%
\caption{}%
\label{subfig.house.c}%
\end{subfigure}\hfill%
\begin{subfigure}{.24\columnwidth}
\includegraphics[width=0.98\textwidth]{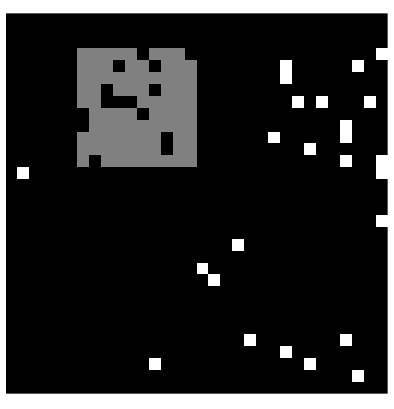}%
\caption{}%
\label{subfig.house.d}%
\end{subfigure}\hfill%
\begin{subfigure}{.24\columnwidth}
\includegraphics[width=\textwidth]{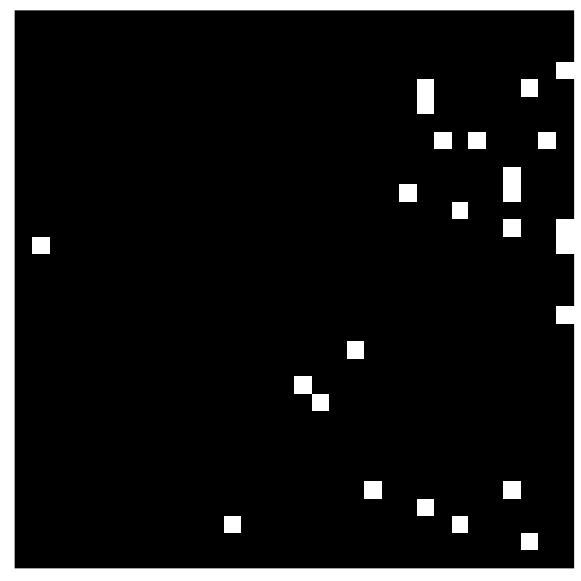}%
\caption{}%
\label{subfig.house.e}%
\end{subfigure}\hfill%
\begin{subfigure}{.24\columnwidth}
\includegraphics[width=\textwidth]{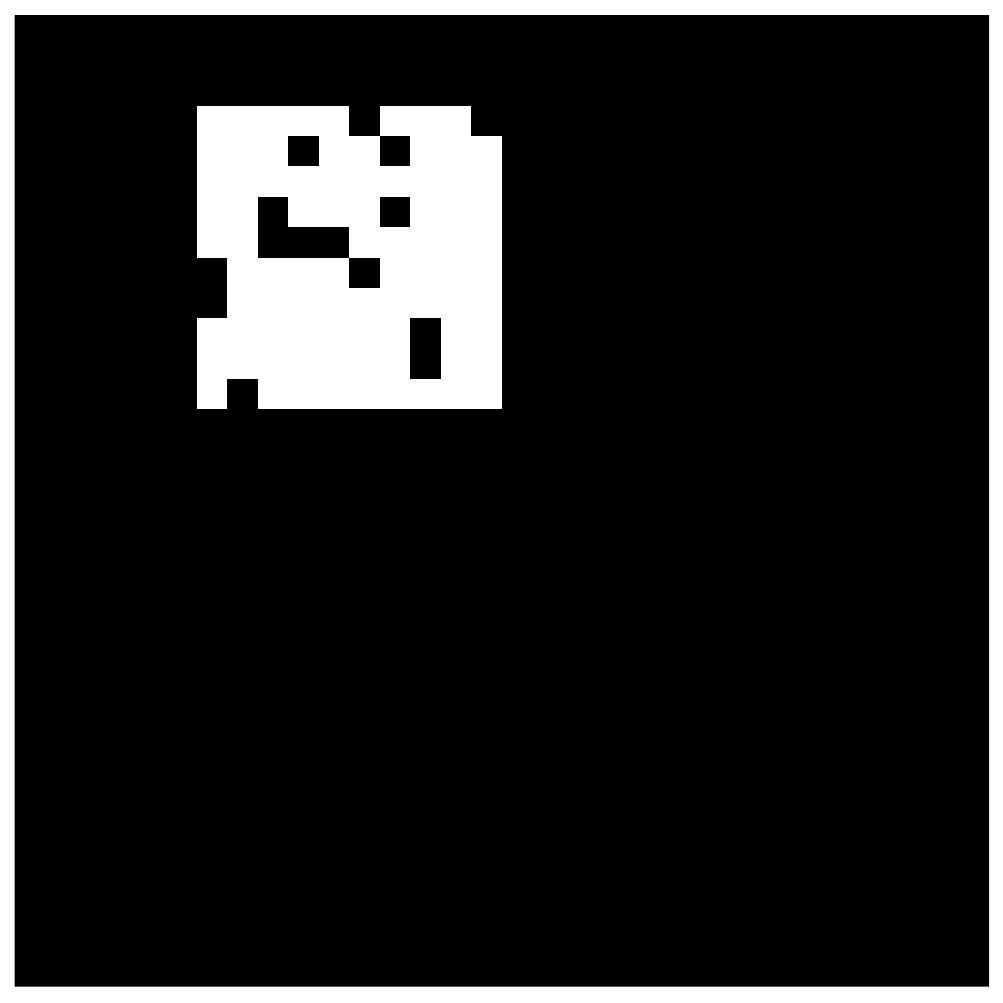}%
\caption{}%
\label{subfig.house.f}%
\end{subfigure}\hfill%
\begin{subfigure}{.24\columnwidth}
\includegraphics[width=\textwidth]{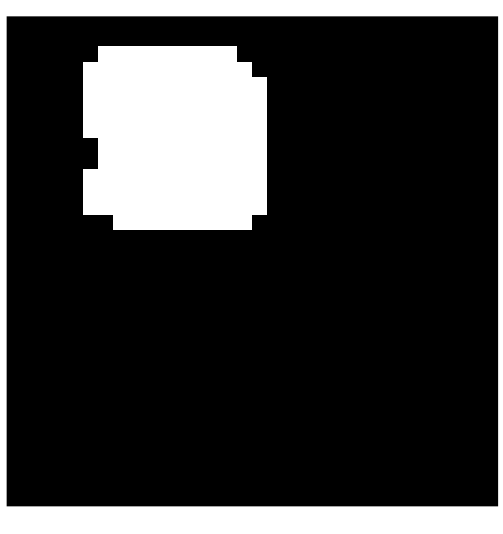}%
\caption{}%
\label{subfig.house.g}%
\end{subfigure}%
\label{Fig.HouseTamperDetection}
\caption{Tamper detection steps on the House standard image. ّFig.\ref{subfig.house.a} is the watermarked image with PSNR=34.38 dB. Fig.\ref{subfig.house.b}, shows the tampeded and JPEG compressed image with QF=50. Fig.\ref{subfig.house.c}, shows the three detected block types (black is the smooth block, gray is the fully destroyed block, white is the rough block). Fig.\ref{subfig.house.d}, shows the conversion of partially destroyed block to fully destroyed block if they have continuty. Fig.\ref{subfig.house.e}, shows the blocks that do not have any role in the continuing tamper detection process. Fig.\ref{subfig.house.f}, shows the continuous partially destroyed blocks and fully destroyed blocks independently. Fig.\ref{subfig.house.f}, shows the tamper deteted blocks after filling the hole and contour blocks (FR=0\% and FA=3.7\%).
 } \label{fig.TamperDetection2}
\end{figure}

\begin{figure*}[!htbp]%
\centering

\begin{subfigure}{0.22\textwidth}

\includegraphics[width=\textwidth]{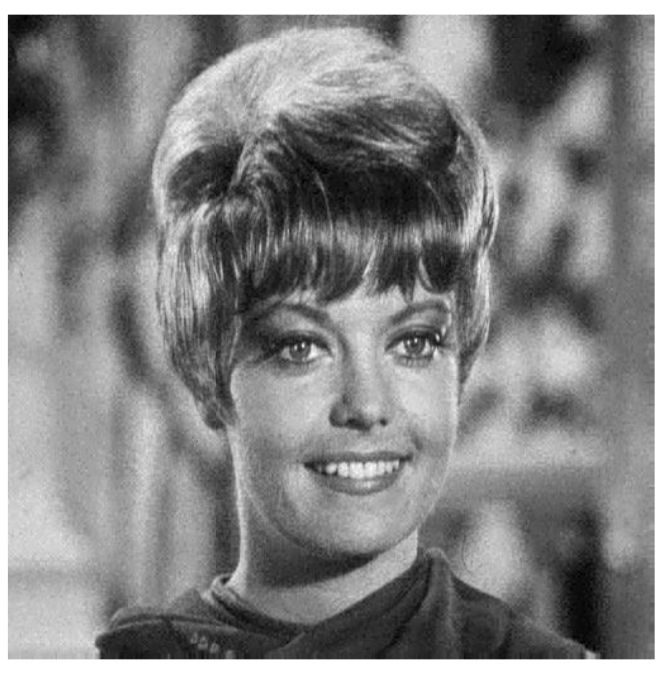}%
\caption{\footnotesize PSNR=34.62 dB}%
\label{subfig.tamperdetection.a}%
\end{subfigure}\hfill%
\begin{subfigure}{.22\textwidth}
\includegraphics[width=\textwidth]{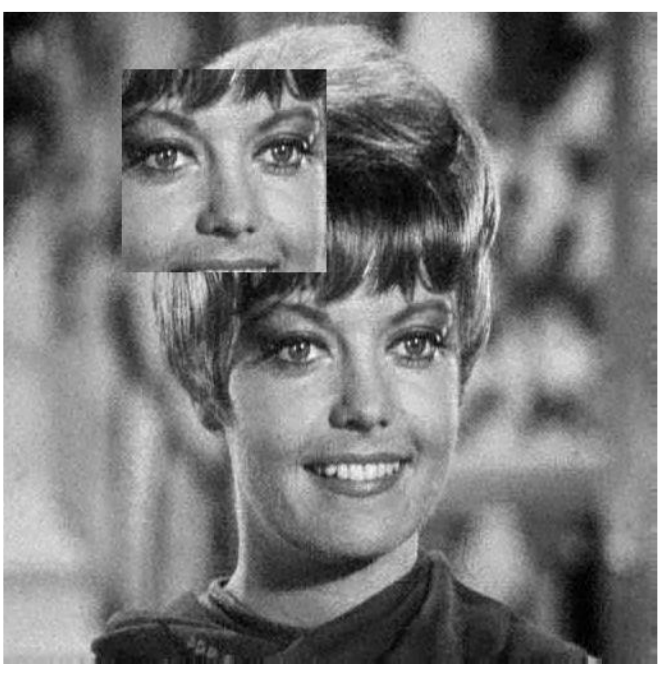}%
\caption{ \footnotesize  }%
\label{subfig.tamperdetection.b}%
\end{subfigure}\hfill%
\begin{subfigure}{.22\textwidth}
\includegraphics[width=\textwidth]{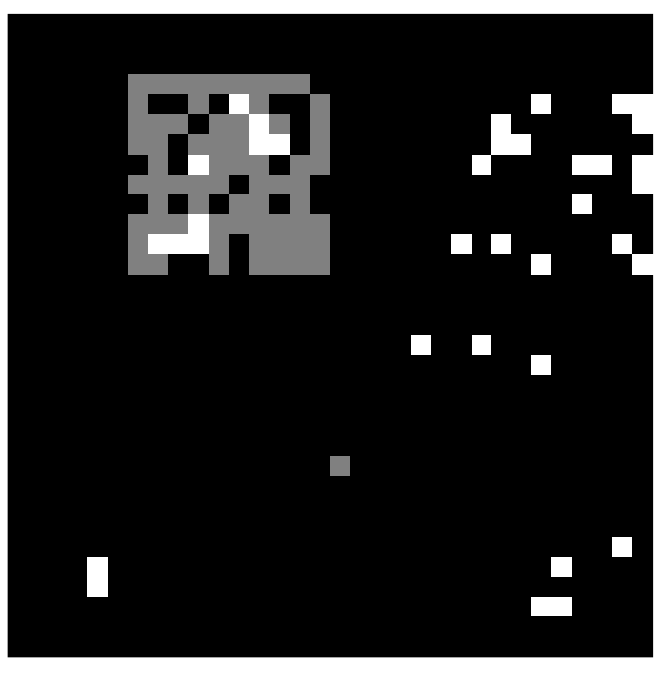}%
\caption{\footnotesize }%
\label{subfig.tamperdetection.c}%
\end{subfigure}\hfill%
\begin{subfigure}{.22\textwidth}
\includegraphics[width=\textwidth]{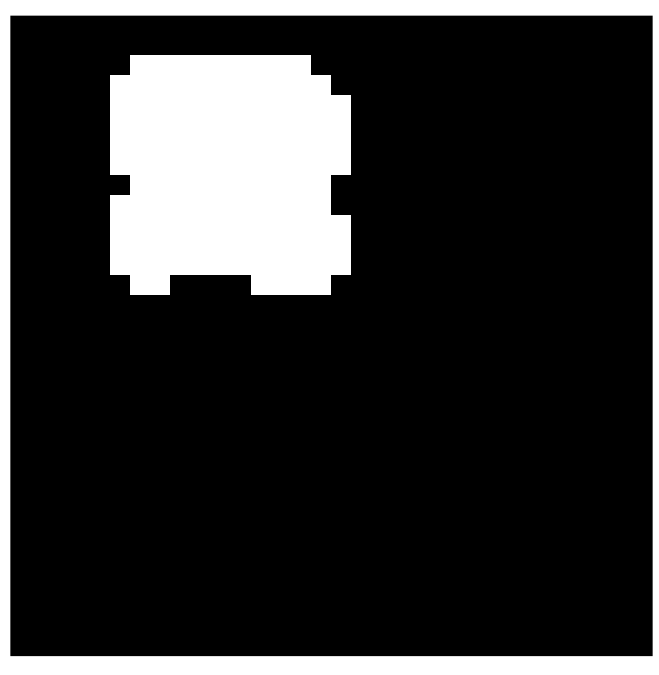}%
\caption{\footnotesize  FR=0\%, FA=3.3\%}%
\label{subfig.tamperdetection.d}%
\end{subfigure}\hfill%
\begin{subfigure}{0.22\textwidth}
\includegraphics[width=\textwidth]{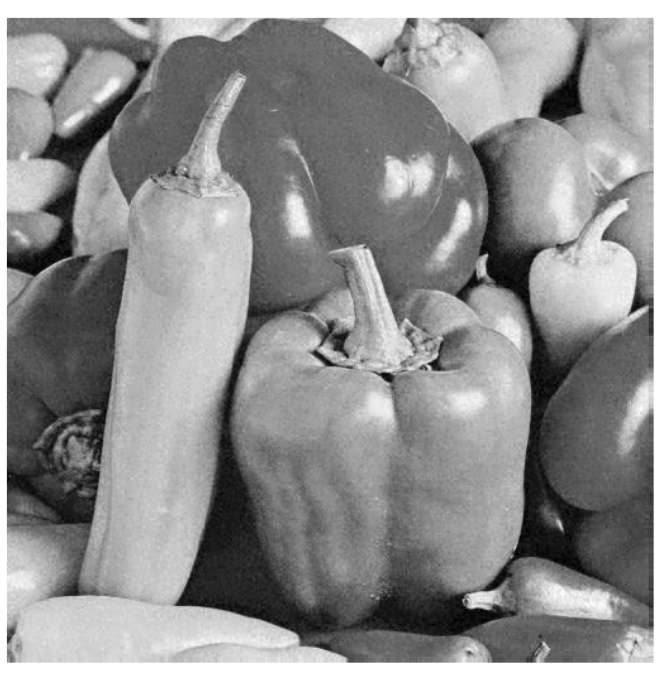}%
\caption{\footnotesize  PSNR=35.20 dB}%
\label{subfig.tamperdetection.e}%
\end{subfigure}\hfill%
\begin{subfigure}{.22\textwidth}
\includegraphics[width=\textwidth]{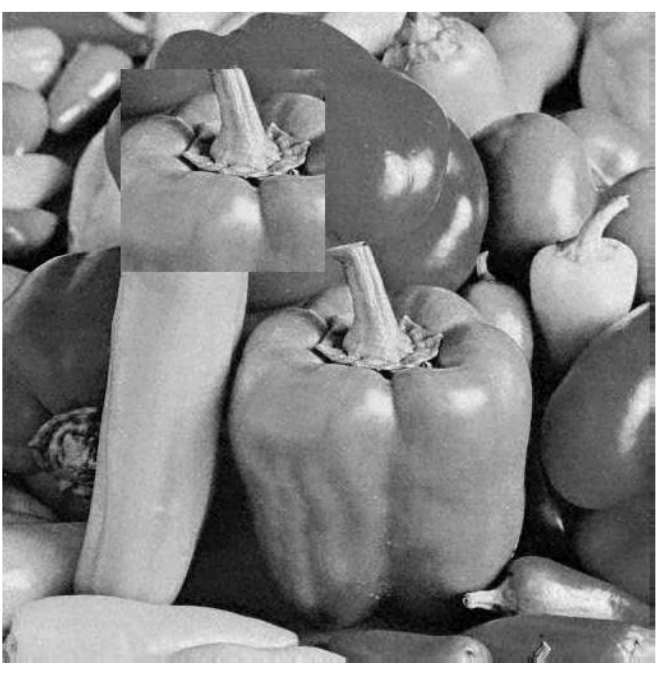}%
\caption{\footnotesize }%
\label{subfig.tamperdetection.f}%
\end{subfigure}\hfill%
\begin{subfigure}{.22\textwidth}
\includegraphics[width=\textwidth]{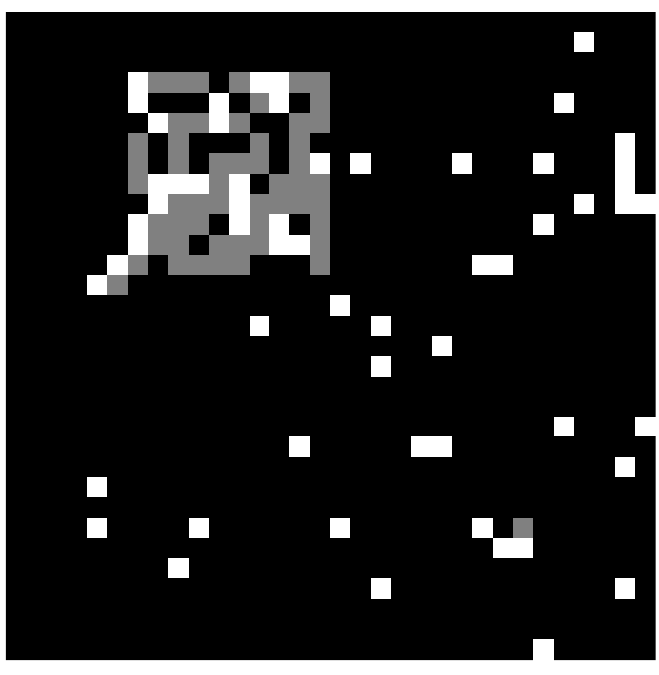}%
\caption{\footnotesize }%
\label{subfig.tamperdetection.g}%
\end{subfigure}\hfill%
\begin{subfigure}{.22\textwidth}
\includegraphics[width=\textwidth]{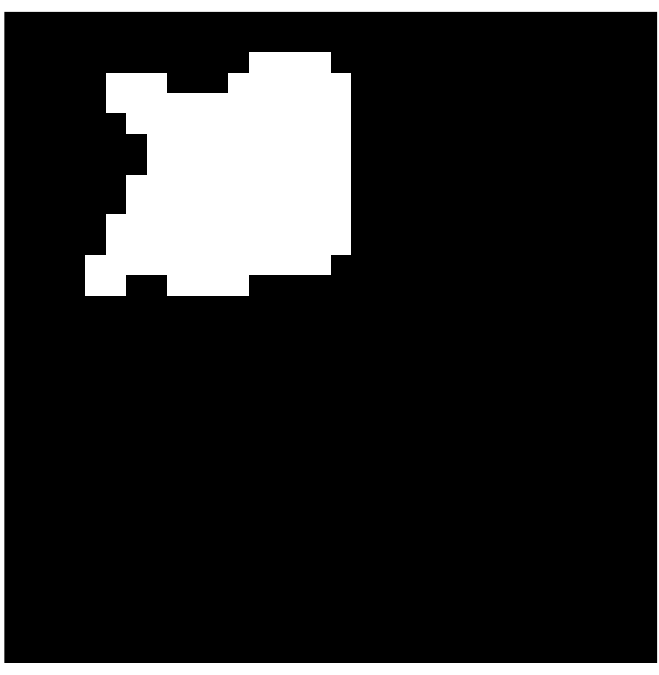}%
\caption{\footnotesize FR=5\%, FA=2.7\%}%
\label{subfig.tamperdetection.h}%
\end{subfigure}\hfill%
\begin{subfigure}{0.22\textwidth}
\includegraphics[width=\textwidth]{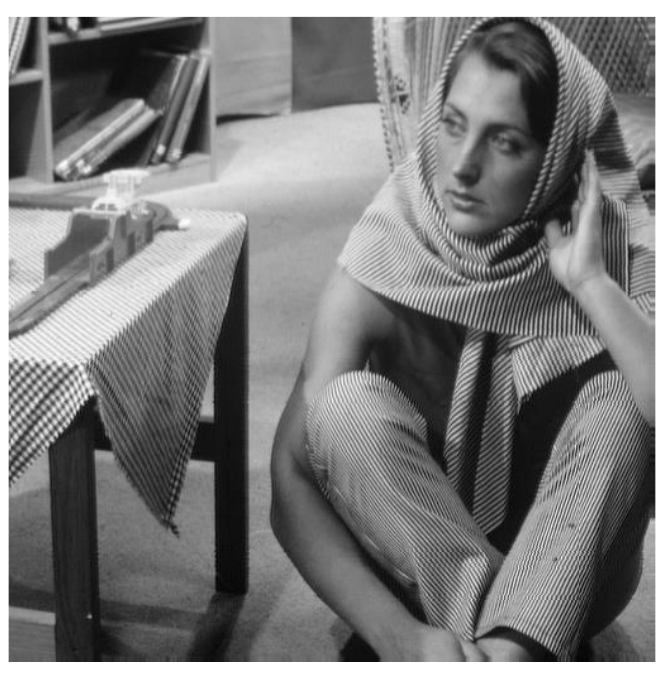}%
\caption{\footnotesize PSNR=34.16 dB}%
\label{subfig.tamperdetection.i}%
\end{subfigure}\hfill%
\begin{subfigure}{.22\textwidth}
\includegraphics[width=\textwidth]{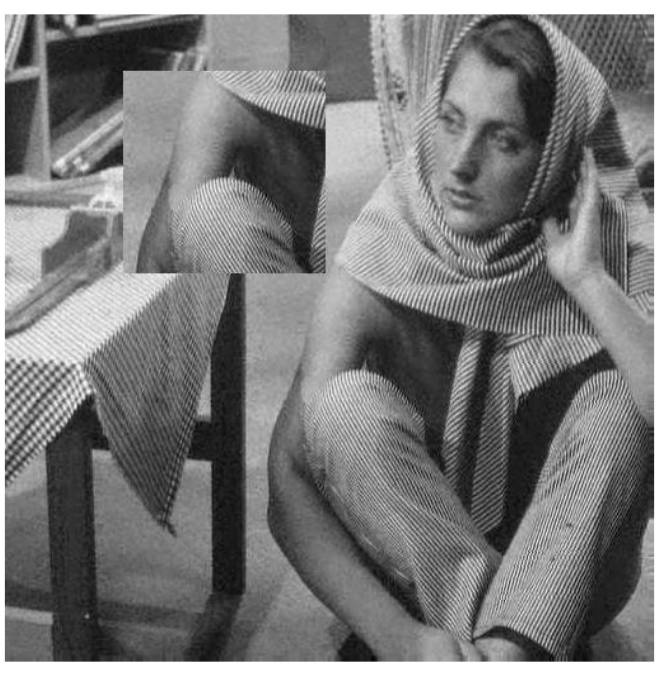}%
\caption{\footnotesize }%
\label{subfig.tamperdetection.j}%
\end{subfigure}\hfill%
\begin{subfigure}{.22\textwidth}
\includegraphics[width=\textwidth]{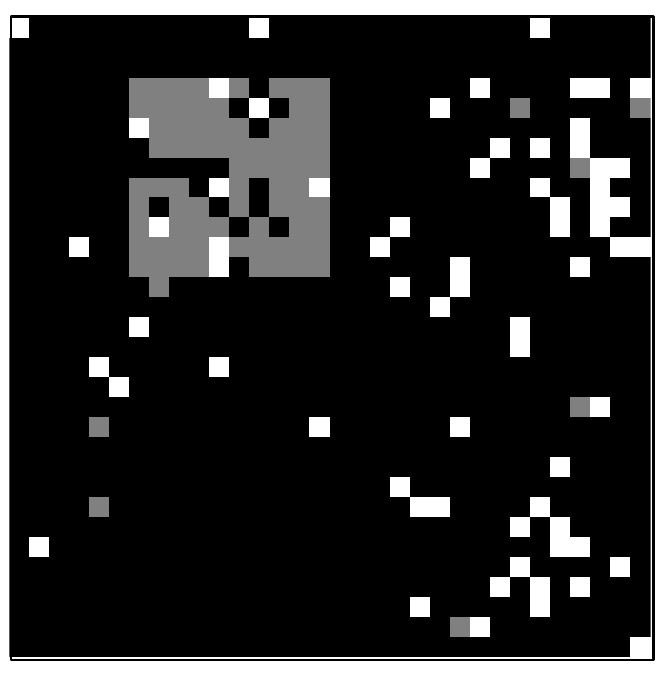}%
\caption{\footnotesize}%
\label{subfig.tamperdetection.k}%
\end{subfigure}\hfill%
\begin{subfigure}{.22\textwidth}
\includegraphics[width=\textwidth]{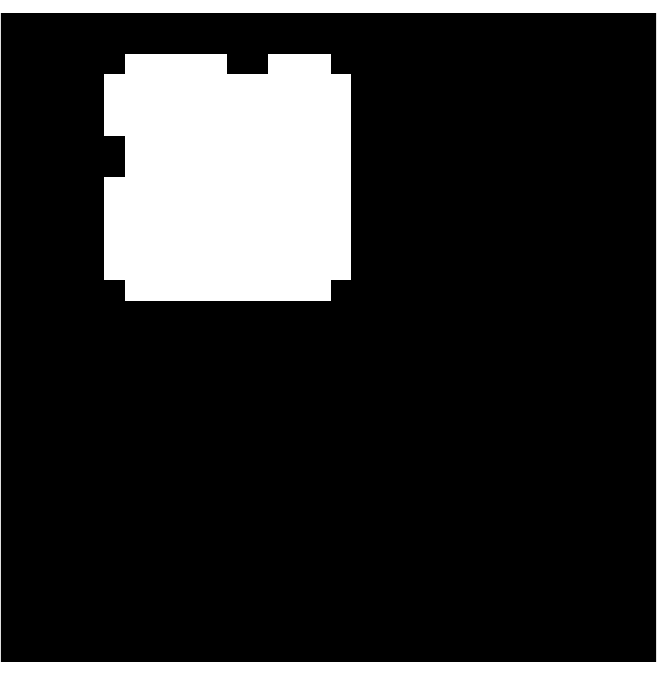}%
\caption{\footnotesize  FR=0\%, FA=3.8\%}%
\label{subfig.tamperdetection.l}%
\end{subfigure}\hfill%
\begin{subfigure}{0.22\textwidth}
\includegraphics[width=\textwidth]{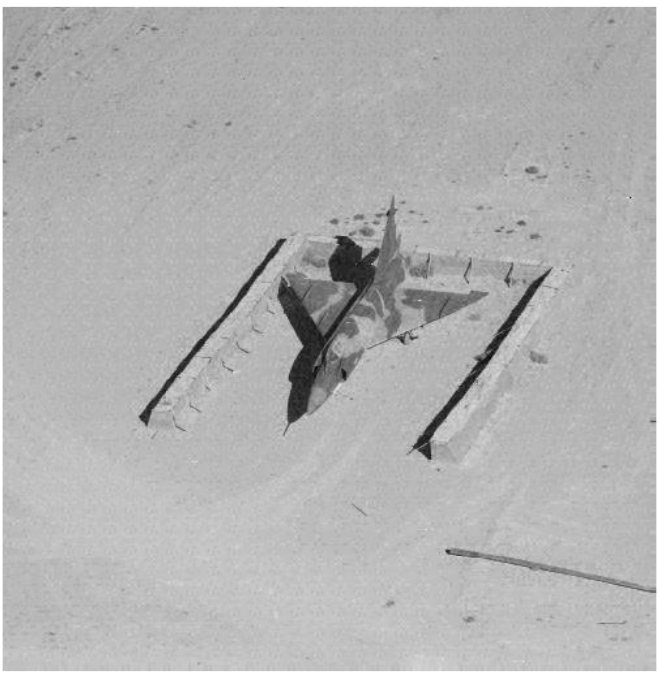}%
\caption{\footnotesize PSNR=36.71 dB}%
\label{subfig.tamperdetection.m}%
\end{subfigure}\hfill%
\begin{subfigure}{.22\textwidth}
\includegraphics[width=\textwidth]{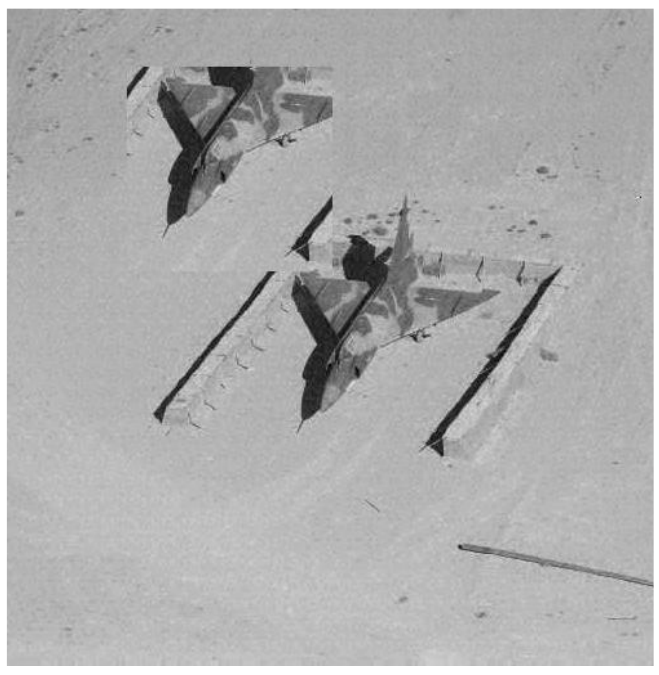}%
\caption{\footnotesize}%
\label{subfig.tamperdetection.n}%
\end{subfigure}\hfill%
\begin{subfigure}{.22\textwidth}
\includegraphics[width=\textwidth]{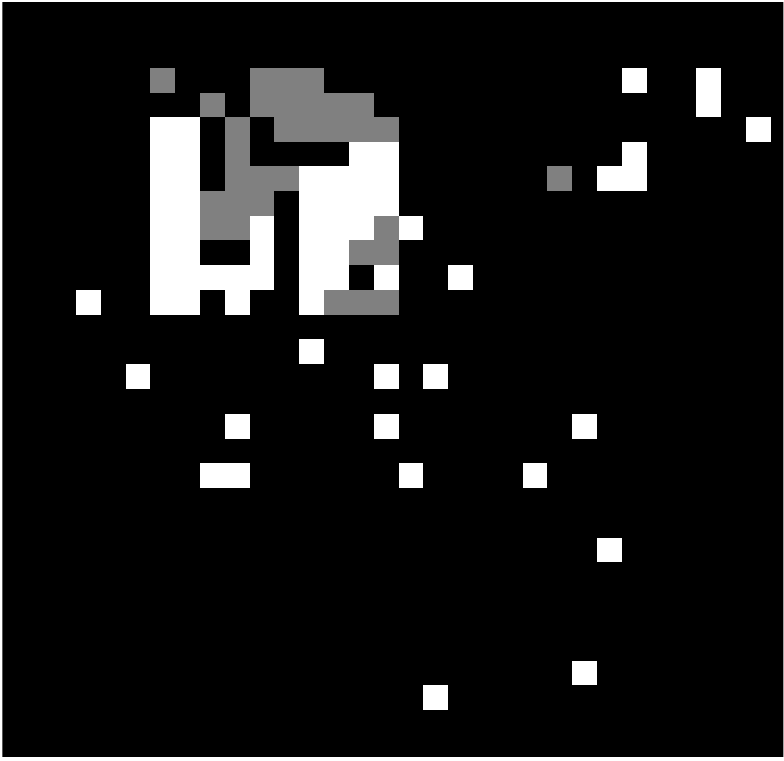}%
\caption{\footnotesize }%
\label{subfig.tamperdetection.o}%
\end{subfigure}\hfill%
\begin{subfigure}{.22\textwidth}
\includegraphics[width=\textwidth]{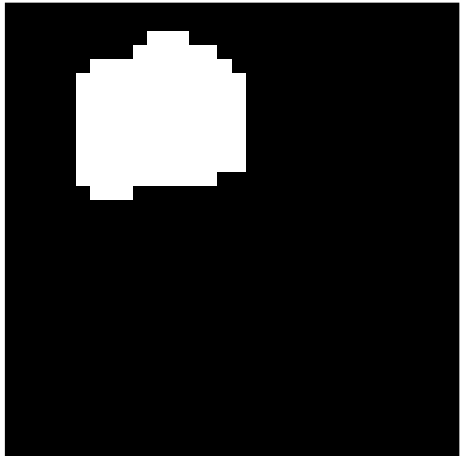}%
\caption{\footnotesize FR=5\%, FA=2.2\%}%
\label{subfig.tamperdetection.p}%
\end{subfigure}\hfill%

\begin{subfigure}{0.22\textwidth}
\includegraphics[width=\textwidth]{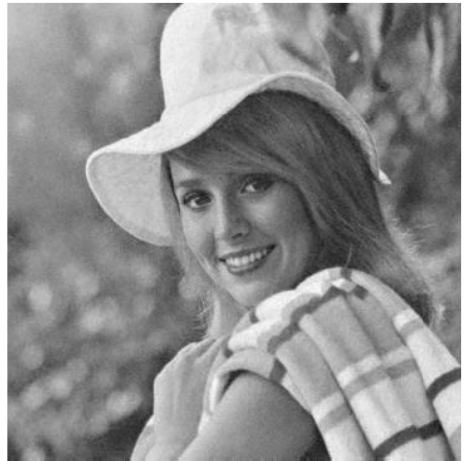}%
\caption{\footnotesize PSNR=35.16 dB}%
\label{subfig.tamperdetection.q}%
\end{subfigure}\hfill%
\begin{subfigure}{.22\textwidth}
\includegraphics[width=\textwidth]{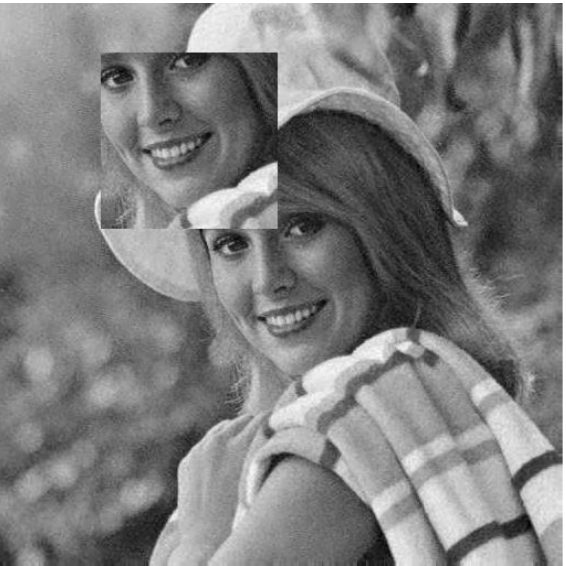}%
\caption{\footnotesize}%
\label{subfig.tamperdetection.r}%
\end{subfigure}\hfill%
\begin{subfigure}{.22\textwidth}
\includegraphics[width=\textwidth]{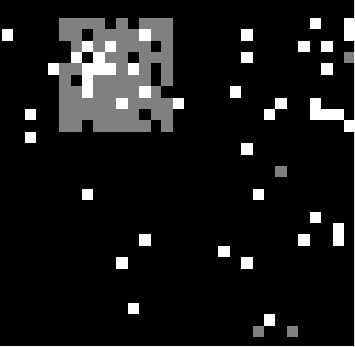}%
\caption{\footnotesize }%
\label{subfig.tamperdetection.s}%
\end{subfigure}\hfill%
\begin{subfigure}{.22\textwidth}
\includegraphics[width=0.98\textwidth]{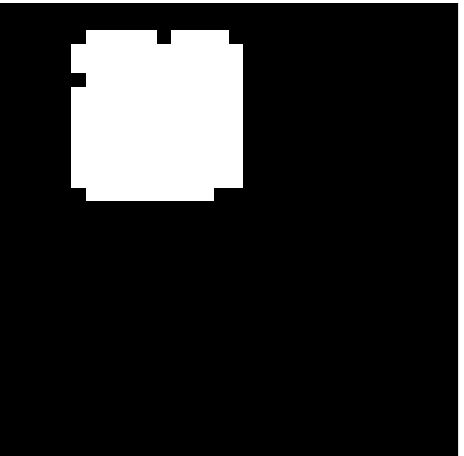}%
\caption{\footnotesize FR=0\%, FA=4\%}%
\label{subfig.tamperdetection.t}%
\end{subfigure}\hfill%

\label{Fig.FiveImageTamperDetection}
\caption{Tamper detection results on some standard images. First column is the watermarked image. Second column is tampered image, which the size of tampered region is considered $160\times160$ and the JPEG compression power is considered QF=50\%. Third column  is the detection of three blocks status (Black is healthful, Gray is fully destroyed and White is partially destroyed). Fourth column is the tampered region.}\label{fig.TamperDetection3}
\end{figure*}


 \begin{figure}[h]
  \centering
  \includegraphics[width=0.9\columnwidth]{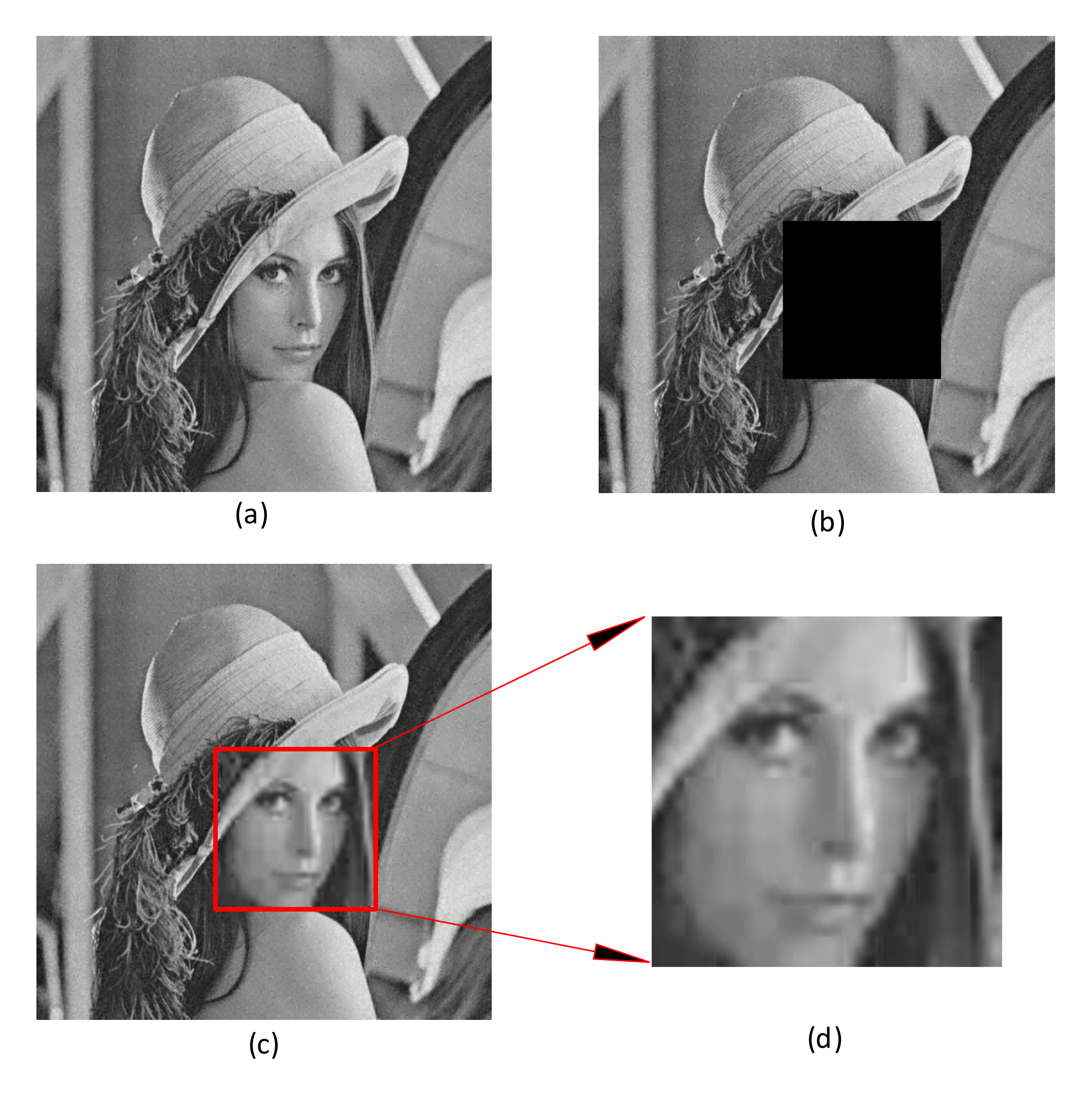}
  \caption{ Fig.\ref{fig.LenaFace}a shows the watermarked image with PSNR 35.20 dB.  Fig.\ref{fig.LenaFace}b shows the tampered region ($176\times 176$) and the JPEG compression power is QF=80\%. Fig.\ref{fig.LenaFace}c is the recovered image with PSNR=31.12 dB. Fig.\ref{fig.LenaFace}d shows the magnified recovered region with PSNR=24.51 dB.}\label{fig.LenaFace}
\end{figure}


 \begin{figure*}[!htbp]
  \centering
  \includegraphics[width=0.9\columnwidth]{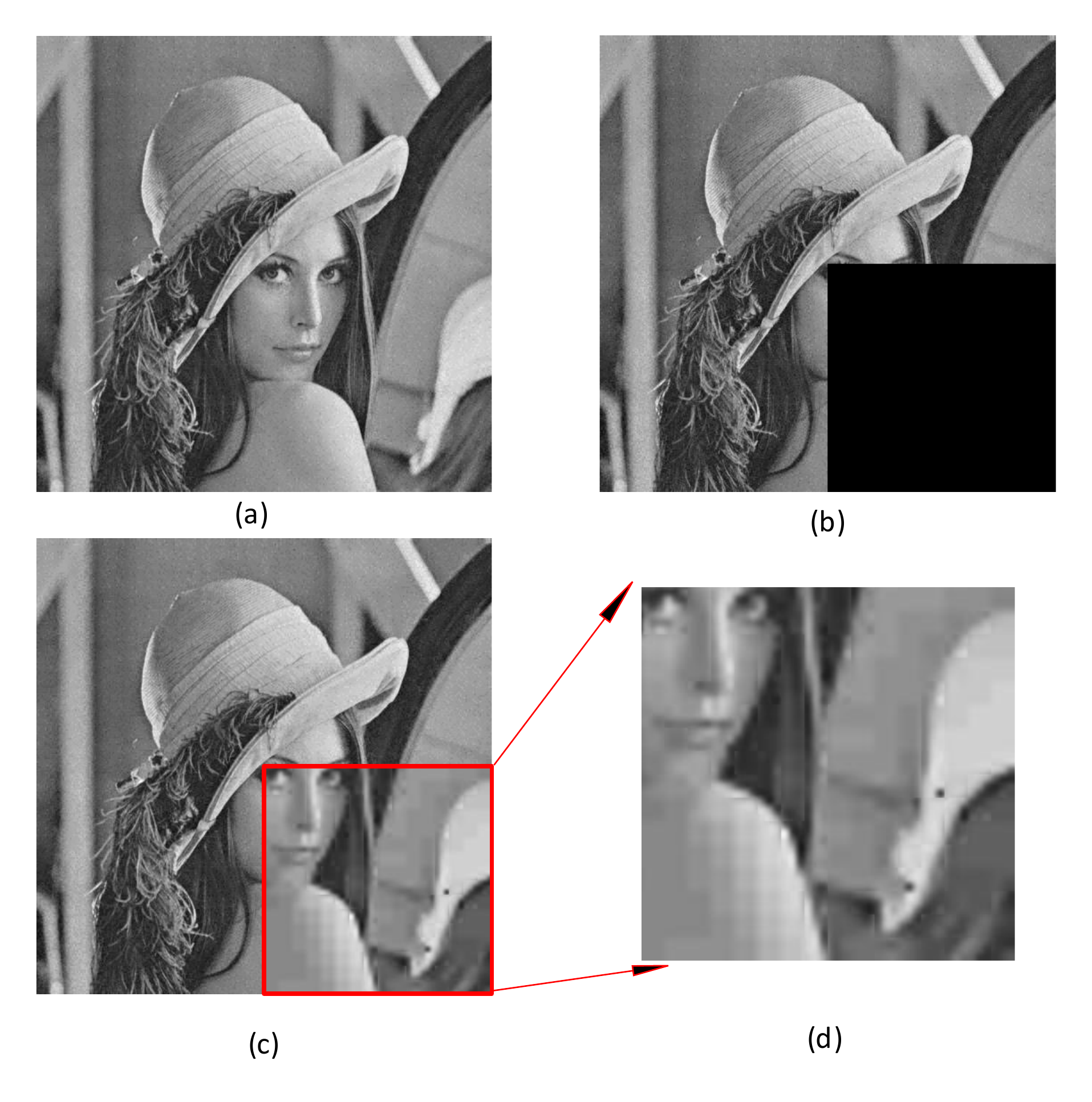}
  \caption{ Fig.\ref{fig.LenaFace}a shows the watermarked image with PSNR 35.20 dB.  Fig.\ref{fig.LenaFace}b shows the tampered region ($256\times 256$) and the JPEG compression power is QF=80\%. Fig.\ref{fig.LenaFace}c is the recovered image with PSNR=30.76 dB. Fig.\ref{fig.LenaFace}d shows the magnified recovered region with PSNR=25.43 dB.}\label{fig.LenaRightDown}
\end{figure*}

\begin{figure*}[!htbp]%
\centering

\begin{subfigure}{0.25\textwidth}

\includegraphics[width=\textwidth]{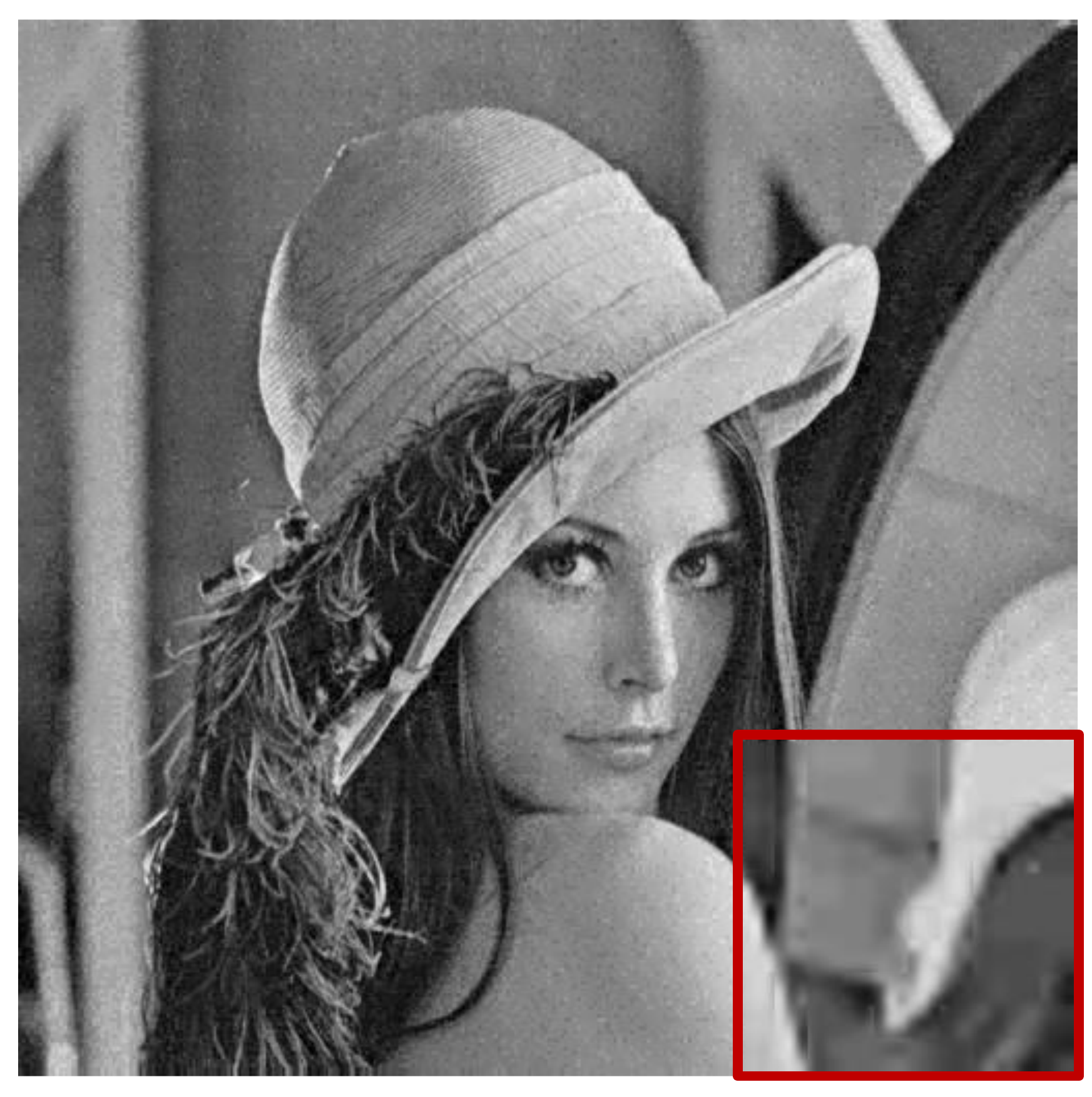}%
\caption{\footnotesize Proposed method}%
\label{subfig.comp.a}%
\end{subfigure}\hfill%
\begin{subfigure}{.25\textwidth}
\includegraphics[width=\textwidth]{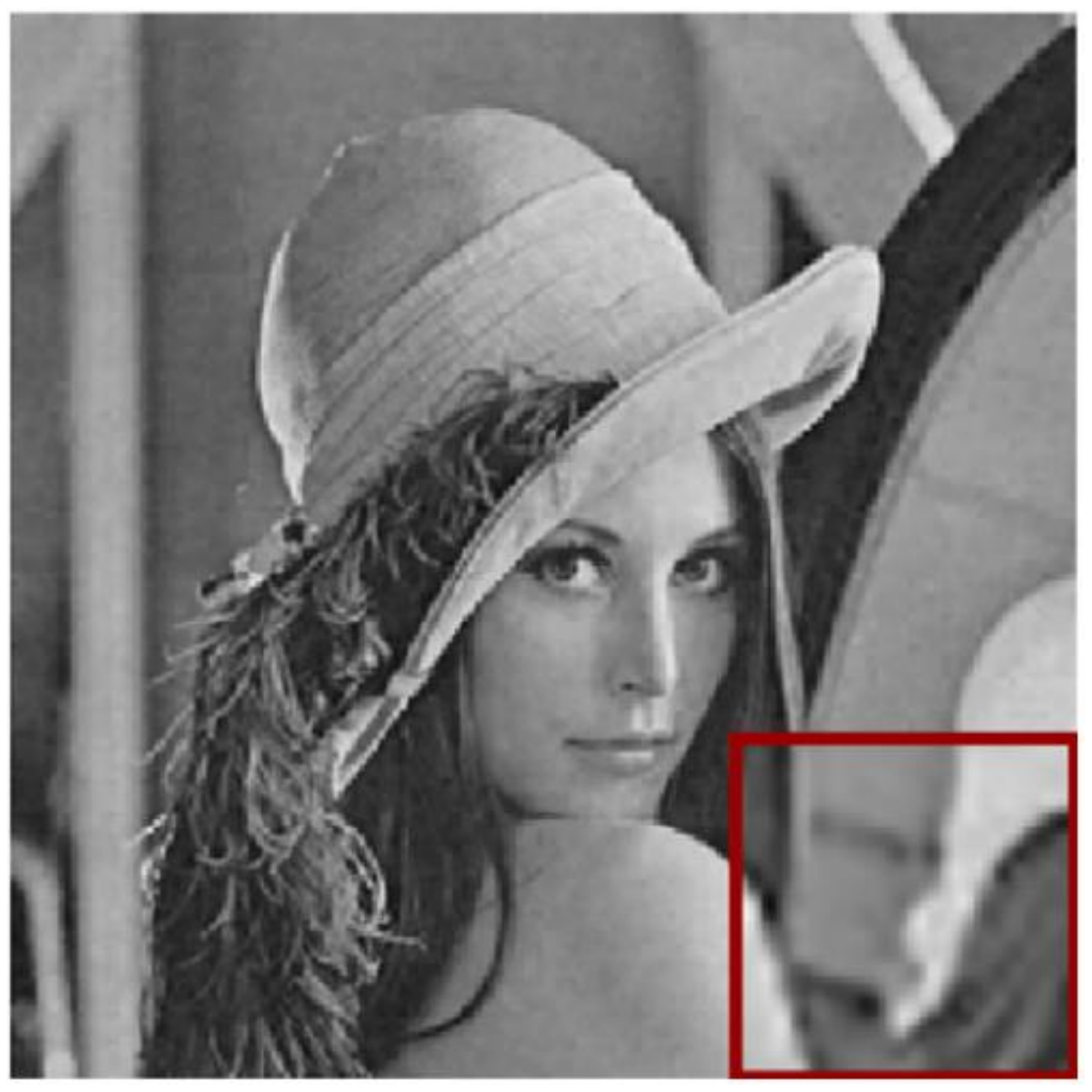}%
\caption{ \footnotesize Method in~\cite{wang2014novel} }%
\label{subfig.comp.b}%
\end{subfigure}\vfill%
\begin{subfigure}{0.25\textwidth}
\includegraphics[width=\textwidth]{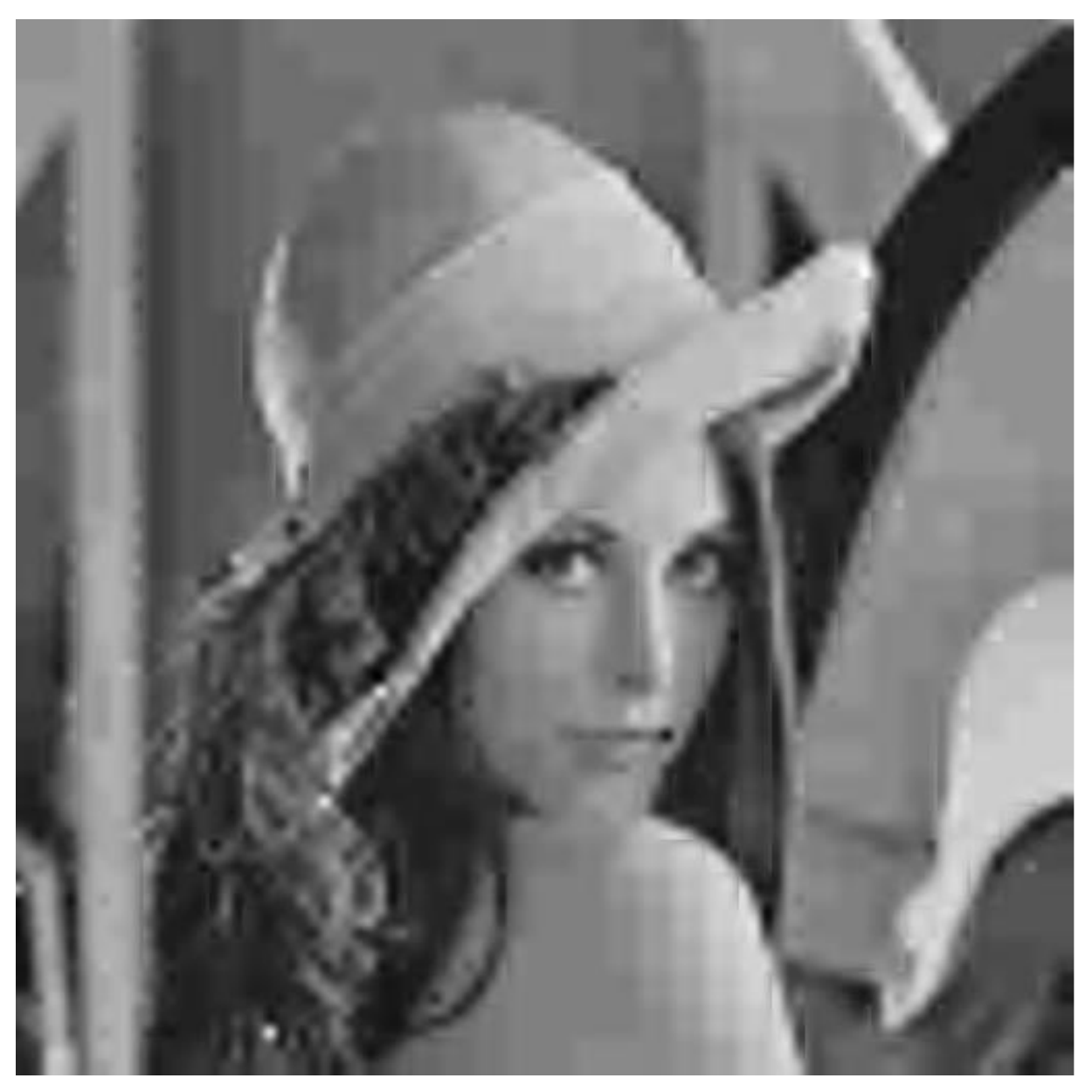}%
\caption{\footnotesize Proposed method}%
\label{subfig.comp.c}%
\end{subfigure}\hfill%
\begin{subfigure}{0.25\textwidth}
\includegraphics[width=\textwidth]{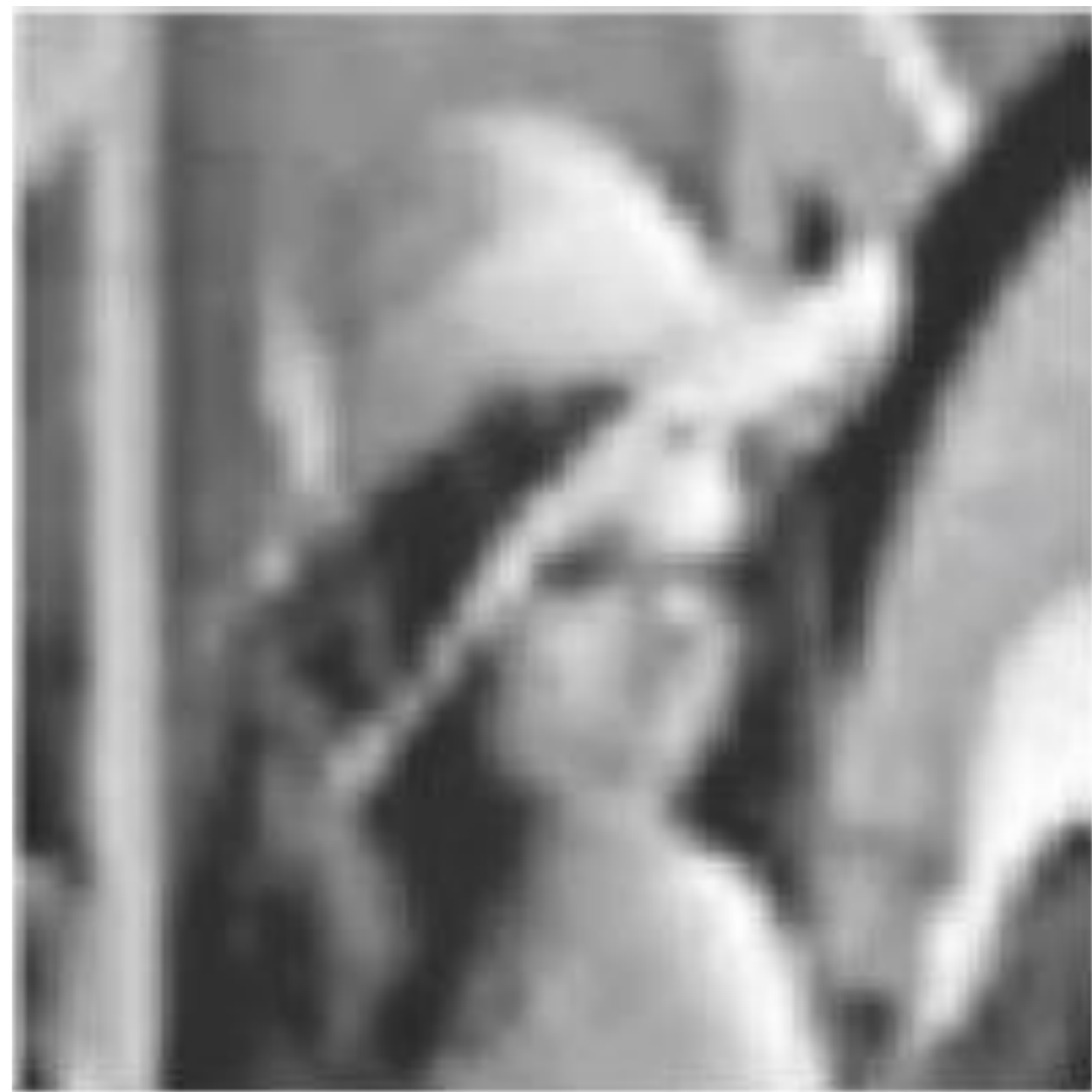}%
\caption{\footnotesize Method in~\cite{li2015semi} }%
\label{subfig.comp.d}%
\end{subfigure}\vfill%
\begin{subfigure}{0.25\textwidth}
\includegraphics[width=\textwidth]{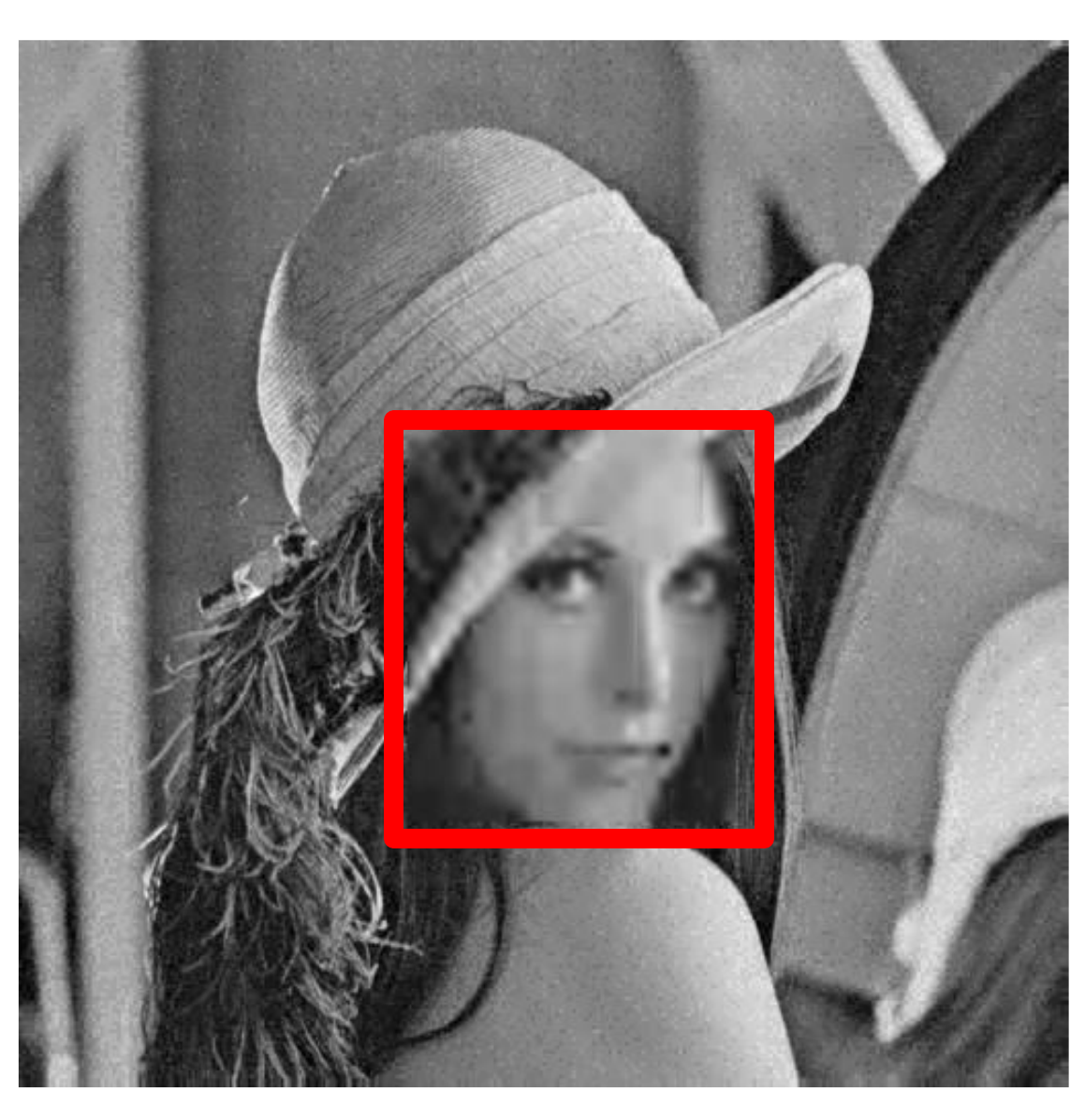}%
\caption{\footnotesize Proposed method}%
\label{subfig.comp.e}%
\end{subfigure}\hfill%
\begin{subfigure}{0.25\textwidth}
\includegraphics[width=\textwidth]{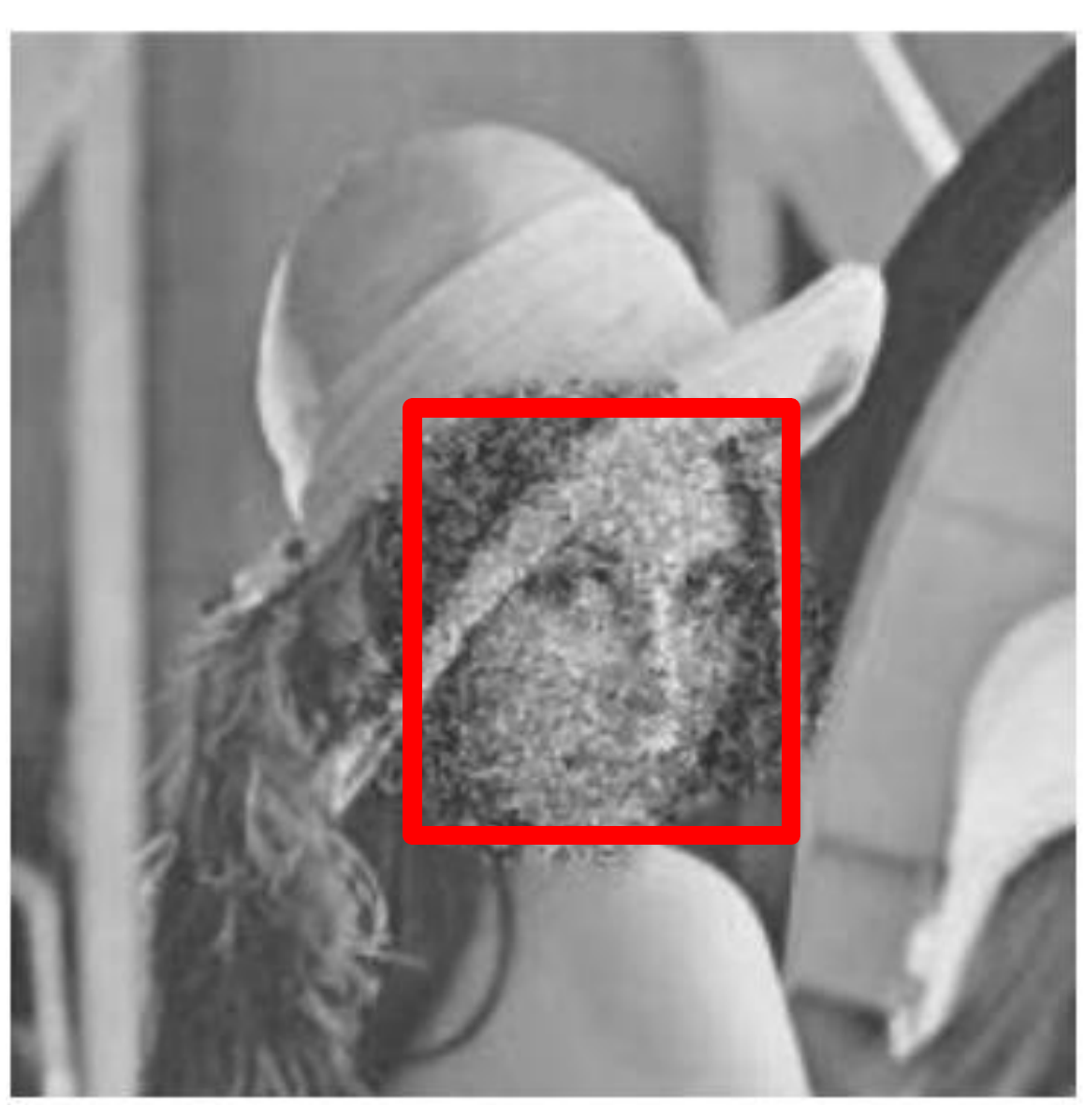}%
\caption{\footnotesize Method in~\cite{phadikar2012novel}}%
\label{subfig.comp.f}%
\end{subfigure}\vfill%
\begin{subfigure}{.40\textwidth}
\includegraphics[width=\textwidth]{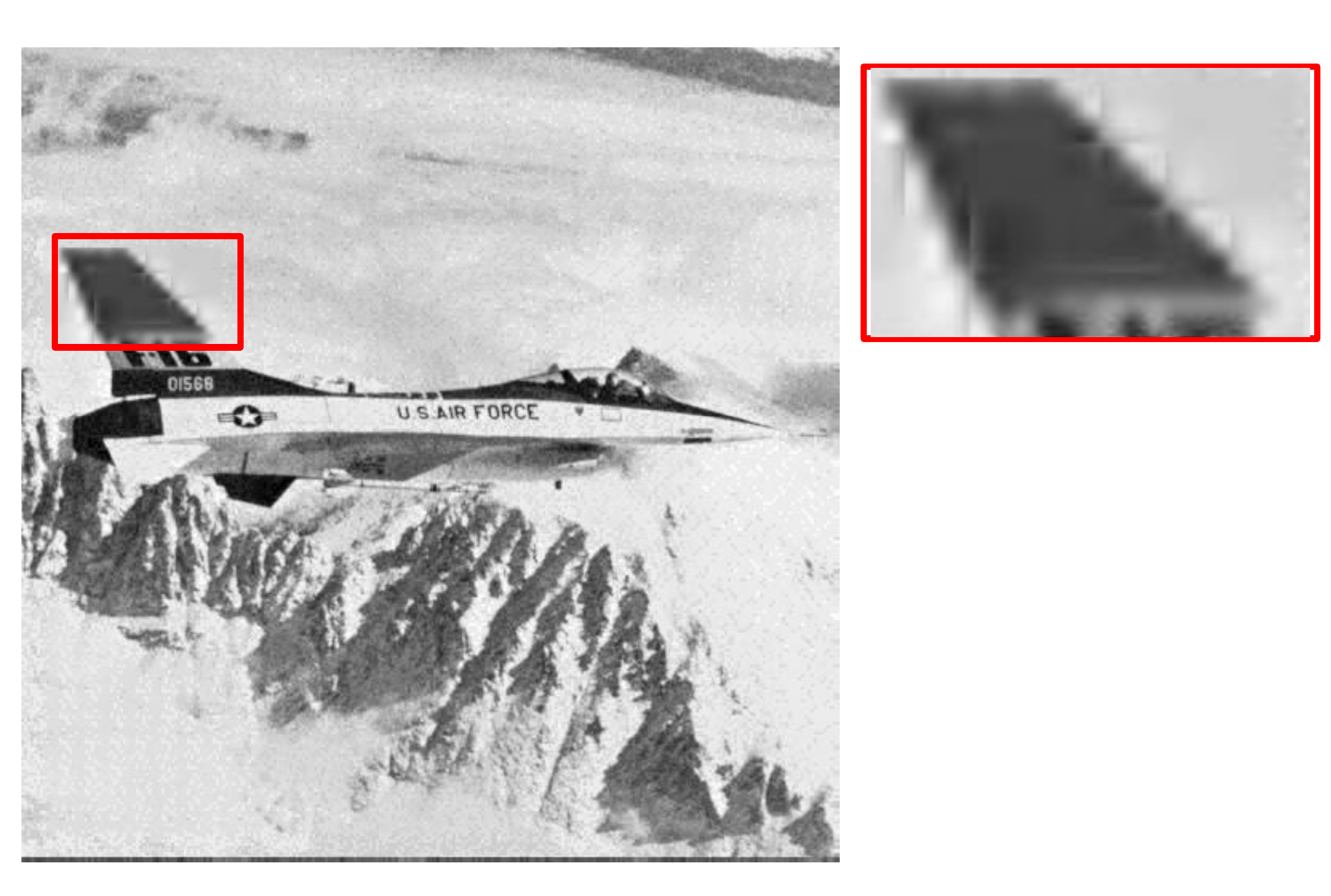}%
\caption{\footnotesize Proposed method}%
\label{subfig.comp.g}%
\end{subfigure}\hfill%
\begin{subfigure}{.40\textwidth}
\includegraphics[width=\textwidth]{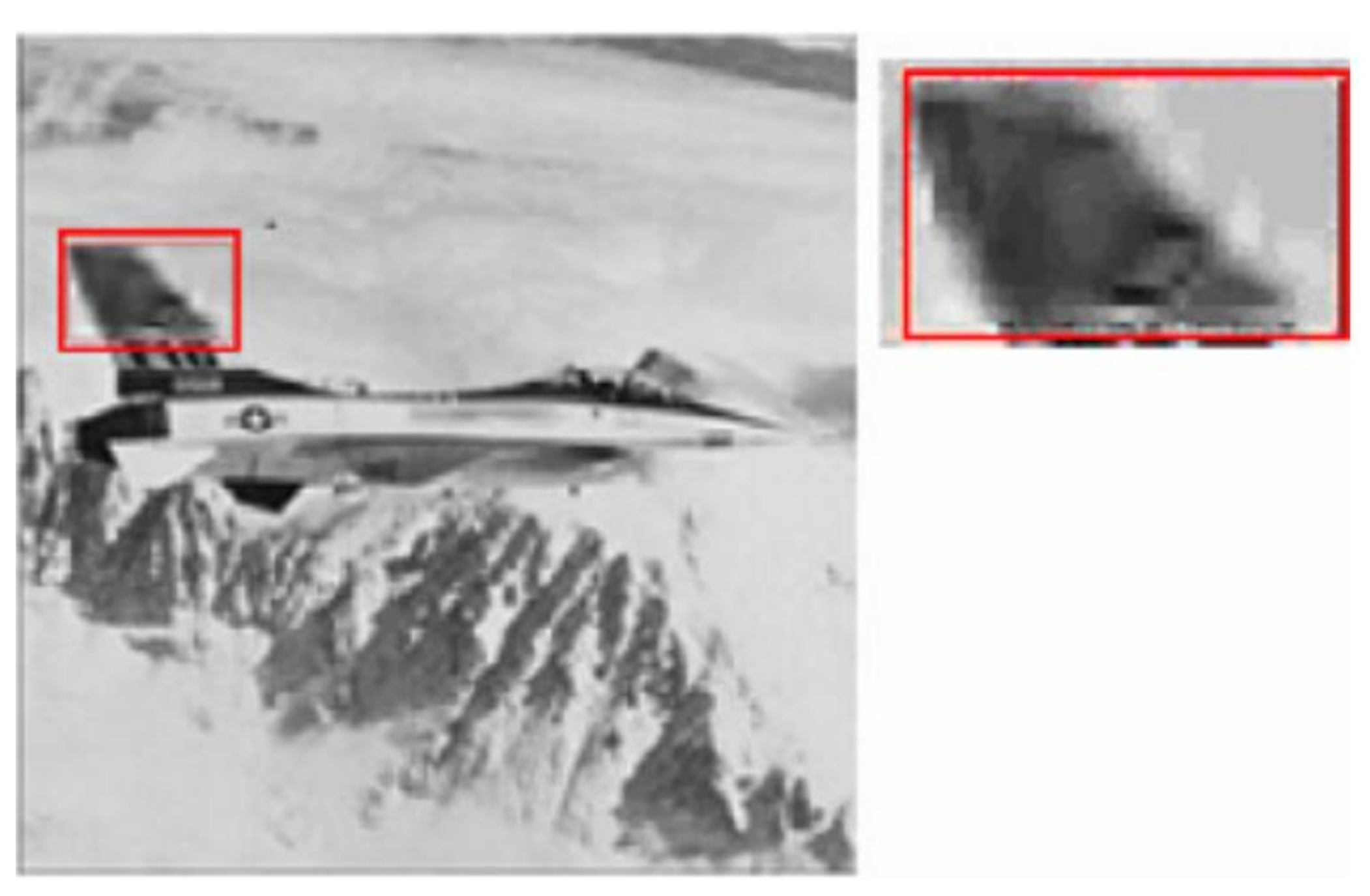}%
\caption{\footnotesize  Method in~\cite{li2014semi}}%
\label{subfig.comp.h}%
\end{subfigure}%

\caption{Visual comparison between the proposed method and methods~\cite{wang2014novel, li2015semi, li2014semi, phadikar2012novel}. All images except for the image in Fig.\ref{subfig.comp.f} are in presence of JPEG compression with QF=80\%. The Fig.\ref{subfig.comp.f} is in presence of JPEG compression with QF=85\%. In Figs.\ref{subfig.comp.a}, \ref{subfig.comp.b}, \ref{subfig.comp.e}, \ref{subfig.comp.f}, \ref{subfig.comp.g} and \ref{subfig.comp.h} the tampered and recovered region are shown using red countour. Figs.\ref{subfig.comp.c} and \ref{subfig.comp.d} are full recovery of image using the embedded information without any tampering.}\label{fig.Compsss}
\end{figure*}

\begin{table*},\caption{Results of proposed method in presense of JPEG compression. Tampered region is a $256\times256$ region in the top-left side of the watermarked images.}\label{tbl.CompJPEG}
\centering

 \begin{tabular}{|l | c| c | c  | c  c c c c c | }
 \hline
 \multicolumn{4}{|c|}{} &  \multicolumn{6}{c|}{JPEG compression QF (Quality Factor)} \\ \hline
 Image & \rotatebox[origin=c]{90}{\thead{PSNR\\Watermarked\\ image }} & \rotatebox[origin=c]{90}{\thead{SSIM\\Watermarked\\ image }} &  & 100 & 95 & 90 & 85 & 80 & 75\\ \hline
 House & 34.38 & 0.89 & \multirow{9}{*}{ \rotatebox[origin=c]{90}{\thead{ PSNR of recovered \\  region }}} & 25.55 & 25.55 & 25.55 & 25.55 & 24.45 &  16.91\\
  Lena & 35.20  & 0.88 & & 26.88 & 26.88 & 26.88 & 26.85 & 26.16 &  24.92\\Zelda & 34.62  & 0.85 &   & 25.47 & 25.47  & 25.47  &  25.47  & 25.43 & 23.92\\ Pepper & 35.20 & 0.88 &   & 27.90 & 27.90 & 27.90 &  27.53 & 27.10 & 17.61\\Woman &  34.16  & 0.90 &  & 22.27 & 22.27 & 22.27 &  22.23 & 21.85 & 20.46\\ Airplan & 36.71  & 0.86 &  & 28.96 & 28.96 & 28.96 &  28.97 & 28.96 & 27.75\\ Elaine &   35.16 & 0.89 & & 27.11 & 27.11 & 27.11 &  27.11 & 22.98 & 23.06\\ Boat & 34.46  & 0.90 &  & 25.65 & 25.65 & 25.65 &  25.63 & 25.30 & 24.28\\ Moon & 35.63  & 0.87 &  & 28.00 & 28.00 & 28.00 &  28.00 & 27.58 & 26.51\\ \hline \textbf{Average}  & \textbf{35.06}  & \textbf{0.88}  &   & \textbf{25.91} & \textbf{25.91} & \textbf{25.91} &  \textbf{25.84} & \textbf{25.06} & \textbf{21.85}\\\hline

 \end{tabular}
\end{table*}

\begin{table*},\caption{Results of proposed method in presense of JPEG2000 compression. Tampered region is a $256\times256$ region in the top-left side of the watermarked images.}\label{tbl.CompJPEG2000}
\centering

 \begin{tabular}{|l | c| c | c  | c  c c c c c | }
 \hline
 \multicolumn{4}{|c|}{} &  \multicolumn{6}{c|}{JPEG2000 compression CR (Compression Ratio)} \\ \hline
 Image & \rotatebox[origin=c]{90}{\thead{PSNR\\Watermarked\\ image }} & \rotatebox[origin=c]{90}{\thead{SSIM\\Watermarked\\ image }} &  & 1 & 2 & 3 & 4 & 5 & 6\\ \hline
 House & 34.38 & 0.89 & \multirow{9}{*}{ \rotatebox[origin=c]{90}{\thead{ PSNR of recovered \\  region }}} & 25.55 & 25.55 & 25.55 & 25.55 & 24.83 &  18.26\\
  Lena & 35.20  & 0.88 & & 25.47 & 25.47 & 25.47 & 25.47 & 25.46 &  23.65\\Zelda & 34.62  & 0.85 &   & 27.90 & 27.90  & 27.90  &  27.90  & 27.91 & 27.71\\ Pepper & 35.20 & 0.88 &   & 22.23 & 22.23 & 22.23 &  22.23 & 22.03 & 20.28\\Woman &  34.16  & 0.90 &  & 22.27 & 22.27 & 22.27 &  22.27 & 18.87 & 11.42\\ Airplan & 36.71  & 0.86 &  & 28.96 & 28.96 & 28.96 &  28.96 & 28.96 & 28.96\\ Elaine &   35.16 & 0.89 & & 27.11 & 27.11 & 27.11 &  26.94 & 24.60 & 13.99\\ Boat & 34.46  & 0.90 &  & 25.65 & 25.65 & 25.65 &  25.51 & 23.23 & 17.70\\ Moon & 35.63  & 0.87 &  & 28.00 & 28.00 & 28.00 &  28.00 & 28.00 & 28.00\\ \hline \textbf{Average}  & \textbf{35.06}  & \textbf{0.88}  &   & \textbf{25.91} & \textbf{25.91} & \textbf{25.91} &  \textbf{25.87} & \textbf{24.88} & \textbf{21.11}\\\hline
 \end{tabular}
\end{table*}

\begin{table*},\caption{Comparison between the quality  of proposed method and some related works.}\label{tbl.CompRelatedWorks}
\centering

 \begin{tabular}{|l | c  c  c  | }
 \hline
 Method & \thead{PSNR \\Watermarked Image } & \thead{SSIM \\Watermarked Image } & \thead{PSNR \\Recovered Region }\\ \hline
  Li 2014~\cite{li2015semi} & 36.00  & 0.81 & 21.00 \\
   Wang 2014~\cite{wang2014novel}  &  37.41 & 0.95 & 21.68\\
     Korus 2015~\cite{korus2014iterative}  & 36.00 & -- &  21.00\\
      Phadikar 2012~\cite{phadikar2012novel}  & 35.00 & 0.86 & 23.46 \\ \hline
      \textbf{Proposed Method}  & \textbf{35.06} & \textbf{0.88} & \textbf{25.06}\\  \hline
  \end{tabular}
\end{table*}

Another visual quality measure that is used in this paper is Structural Similarity (SSIM), which the structure of the image is considered in it. The values of SSIM measure is in the range [0,1], which 0 is the minimum similarity and 1 is the maximum similarity.

Dividing the original image into $16\times 16$ blocks and categorization of them into smooth, normal and rough blocks according to the standard deviation of each block is shown in Fig.~\ref{fig.ResultsCategorization}. Two thresholds $Th_1$ and $Th_2$ are set to 0.1 and 0.3, respectively, for categorization of the normalized standard deviation values. The normalized values in range [0,1), [0.1,0.3) and [0.3,1.0] are smooth, normal and rough blocks, respectively.

The  Figs.~\ref{fig.TamperDetection1},~\ref{fig.TamperDetection2}~and~\ref{fig.TamperDetection3} are examples of tamper detection operations. In order to enhance understanding, more steps of tamper detection steps are shown in Figs.~\ref{fig.TamperDetection1} and~\ref{fig.TamperDetection2}. The Fig.~\ref{fig.TamperDetection3} shows the summary of the tamper detection steps for some of the standard images. In these examples, the size of the tampered region is considered $160\times160$ (equal to 100 blocks) and the JPEG compression power that is applied to the watermarked image is QF=50\%. In Fig.\ref{subfig.lena.c}, the three detected block types (black is the smooth block, gray is the fully destroyed block, white is the rough block) are shown. The Fig.\ref{subfig.lena.d}, shows the conversion of the partially destroyed block to fully destroyed block if they have continuity. This continuity between the partially destroyed block and the fully destroyed blocks can be either directly or through some intermediate partially destroyed blocks. The Fig.\ref{subfig.lena.e}, shows the blocks that do not have any role in the continuing of tamper detection process. Separation of partially destroyed blocks and fully destroyed blocks has enabled us to remove some not tampered blocks in the later steps of tamper detection. This step does not exist in the method~\cite{li2015semi} because in that paper is not any difference between partially destroyed block and fully destroyed block.

 The Fig.~\ref{subfig.lena.f}, shows the combined continuous partially destroyed blocks and fully destroyed blocks independently. And finally, the Fig.~\ref{subfig.lena.f} shows the tamper detected blocks after filling the hole and contour blocks. A $3\times3$ neighborhood kernel is used for the filling operation. In the Fig.~\ref{subfig.lena.f} the healthfull blocks (black color) who have more than two destroyed blocks (white color), are considered as destroyed blocks. The average value of tamper localization power, for the Figs.~\ref{fig.TamperDetection1},~\ref{fig.TamperDetection2}~and~\ref{fig.TamperDetection3}, in terms of false rejection (FR) and false acceptance (FA) is  1.4\% and 3.2\%, respectively. These measures are reported near to zero in method~\cite{li2015semi}.

Recovery results for Lena image are shown in Figs.~\ref{fig.LenaFace} and~\ref{fig.LenaRightDown}. Lena image is a good standard image that is composed of three regions (smooth, normal and rough). As shown in these figures, the smooth region (such as the body) is recovered with lower quality and the rough region (such as edges) is recovered with higher quality. The reason for this difference in the quality of regions is the difference in the amount of extracted and used information bits for the blocks of each region. The visual comparison between the proposed method and methods~\cite{wang2014novel, li2015semi, li2014semi, phadikar2012novel} is shown in Fig.~\ref{fig.Compsss}. In these comparisons, the same region of the image is tampered and recovered. Also, the JPEG compression ratio is equal in the all of them. Due to the Fig.~\ref{fig.SufficientDistance} and the distance between the dependent blocks,
the proposed method is able to detect and recover a square $256\times 256$ region of the image in the best case.

In Tables.~\ref{tbl.CompJPEG}~and~\ref{tbl.CompJPEG2000},  a $256\times256$ region on the top-left side of the watermarked images is tampered and recovered in presence of a different ratio of JPEG compression and JPEG2000 compression, respectively. The quality factor (QF) of JPEG compression in these experiments are 100, 95, 90, 85, 80 and 75. Also, the compression ratio (CR) of JPEG2000 compression is represented by 1, 2, 3, 4, 5 and 6, which CR=2  implies that the output image size is half of the input image size or less. As seen from these tables the average PSNR value for the recovered region is 25.06 dB for JPEG compression with QF=80\% and also is  24.88 dB for JPEG2000 compression with CR=5, which these results are high PSNR values. Applying the cubic interpolation on the reformed sub-blocks values, which is shown in Fig.\ref{fig.ExpandBlk}, makes the proposed algorithm to be efficient in recovery of tampered regions in comparison with the state-of-the-art methods.

Also, the PSNR value after tamper detection and recovery on the 70 miscellaneous standard image from the USC-SIPI image database~\cite{USC} is calculated. In this experiment, a random square portion ($100\times100$) is filled with black gray level and then is detected and recovered. Also, this experiment is done in presence of JPEG compression with QF=80\%. Also, The PSNR Value for the recovered portion of all 70 images is calculated. The mean and variance values of the PSNR are 24.27 dB and 27.35. Some image results are near 15 dB (not acceptable) because the proposed method is reliable in presence of JPEG compression with QF=85\%-100\%. The False Rejection and False Acceptance values of the proposed method are 2.5 and 2.8 percent in the average state.

\section{Conclusions}\label{lbl.conclusions}
 In this paper, a new semi-fragile watermarking algorithm is proposed for tamper detection and recovery. Three main ideas that are proposed in this paper are region attention attitude, tamper localization method and the usage of cubic interpolation method in the recovery phase for solving the blocking problem. Region attention is defined as the extracting fewer information bits for the smooth region of image and more information bits for normal and rough regions of the image. Also, in order to provide more imperceptibility, the information bits extracted for each type of blocks are embedded in a dependent far block with the same type.
  Two side circular block dependency is proposed for increasing the power of tamper localization and solving the weakness of the pairwise dependency and the one side circular block dependency.
 
  In order to recover the tampered regions, the average values of sub-blocks are extracted and used by cubic interpolation. So, a good combination is provided between sub-blocks and blocks. The PSNR measure for watermarked image and recovered region are 35.06 dB and 25.06 dB, respectively. Also, localization power based on the false rejection (FR) and false acceptance (FA) measures are 1.2 \% and 3.4 \%, respectively in the average state.
 
  Usage of halftone image as the extracted information bits for each region can be considered as a future work. One of the most challenges in the usage of halftone image with the proposed framework is the adaptation and mixing the zero and one bits of halftone image in the neighbor blocks.
 
 As seen from the visual results, the original image is divided into $16\time16$ non-overlapping blocks and each smooth, normal and rough block is divided into 1, 4 and 9 fix sub-blocks, respectively. So, as another future work, the quadtree or Q-tree structure can be considered for dividing each high texture block into more sub-blocks in front of fix sub-blocking.

\section*{References}


\begin{thebibliography}{99}
\bibitem{cox2007digital}I. Cox, M. Miller, J. Bloom, J. Fridrich, T. Kalker, Digital watermarking and steganography, Morgan Kaufmann, 2007.\\

\bibitem{nguyen2016reversible}T.-S. Nguyen, C.-C. Chang, X.-Q. Yang, A reversible image authentication scheme based on fragile watermarking in discrete wavelet transform domain, AEU-International J. Electron. Commun. 70 (2016) 1055–1061. doi:10.1016/j.aeue.2016.05.003.\\

\bibitem{yu2015new}M. Yu, J. Wang, G. Jiang, Z. Peng, F. Shao, T. Luo, New fragile watermarking method for stereo image authentication with localization and recovery, AEU-International J. Electron. Commun. 69 (2015) 361–370.\\

\bibitem{ghosal2014binomial}S.K. Ghosal, J.K. Mandal, Binomial transform based fragile watermarking for image authentication, J. Inf. Secur. Appl. 19 (2014) 272–281.\\

\bibitem{chen2014chaos}F. Chen, H. He, H.-M. Tai, H. Wang, Chaos-based self-embedding fragile watermarking with flexible watermark payload, Multimed. Tools Appl. 72 (2014) 41–56.\\

\bibitem{qin2016fragile}C. Qin, P. Ji, J. Wang, C.-C. Chang, Fragile image watermarking scheme based on VQ index sharing and self-embedding, Multimed. Tools Appl. 76 (2017) 2267–2287.\\

\bibitem{taherinia2010new}A.H. Taherinia, M. Jamzad, A new spread spectrum watermarking method using two levels DCT, Int. J. Electron. Secur. Digit. Forensics. 3 (2010) 1–26.\\

\bibitem{soleymani2015robust}S.H. Soleymani, A.H. Taherinia, Robust image watermarking based on ICA-DCT and noise augmentation technique, in: Comput. Knowl. Eng. (ICCKE), 2015 5th Int. Conf., IEEE, 2015: pp. 18–23.\\

\bibitem{soleymani2016double}S.H. Soleymani, A.H. Taherinia, Double expanding robust image watermarking based on Spread Spectrum technique and BCH coding, Multimed. Tools Appl. (2016) 1–19. doi:10.1007/s11042-016-3734-2.\\

\bibitem{preda2013semi}R.O. Preda, Semi-fragile watermarking for image authentication with sensitive tamper localization in the wavelet domain, Measurement. 46 (2013) 367–373.\\

\bibitem{shi2016region}H. Shi, M. Li, C. Guo, R. Tan, A region-adaptive semi-fragile dual watermarking scheme, Multimed. Tools Appl. 75 (2016) 465–495.\\

\bibitem{huo2014semi}Y. Huo, H. He, F. Chen, A semi-fragile image watermarking algorithm with two-stage detection, Multimed. Tools Appl. 72 (2014) 123–149.\\

\bibitem{chamlawi2010digital}R. Chamlawi, A. Khan, Digital image authentication and recovery: employing integer transform based information embedding and extraction, Inf. Sci. (Ny). 180 (2010) 4909–4928. doi:10.1016/j.ins.2010.08.039.\\

\bibitem{ullah2013dual}R. Ullah, A. Khan, A.S. Malik, Dual-purpose semi-fragile watermark: Authentication and recovery of digital images, Comput. Electr. Eng. 39 (2013) 2019–2030. doi:10.1016/j.compeleceng.2013.04.024.\\

\bibitem{wang2011novel}H. Wang, A.T.S. Ho, X. Zhao, A novel fast self-restoration semi-fragile watermarking algorithm for image content authentication resistant to JPEG compression, in: Int. Work. Digit. Watermarking, Springer, 2011: pp. 72–85.\\

\bibitem{wang2014novel}H. Wang, A.T.S. Ho, S. Li, A novel image restoration scheme based on structured side information and its application to image watermarking, Signal Process. Image Commun. 29 (2014) 773–787. doi:10.1016/j.image.2014.05.001.\\

\bibitem{li2015semi}C. Li, A. Zhang, Z. Liu, L. Liao, D. Huang, Semi-fragile self-recoverable watermarking algorithm based on wavelet group quantization and double authentication, Multimed. Tools Appl. 74 (2015) 10581–10604. doi:10.1007/s11042-014-2188-7.\\

\bibitem{li2014semi}Y. Li, L. Du, Semi-fragile watermarking for image tamper localization and self-recovery, in: Secur. Pattern Anal. Cybern. (SPAC), 2014 Int. Conf., IEEE, 2014: pp. 328–333.\\

\bibitem{korus2014iterative}P. Korus, J. Białas, A. Dziech, Iterative filtering for semi-fragile self-recovery, in: 2014 IEEE Int. Work. Inf. Forensics Secur., IEEE, 2014: pp. 36–41. doi:10.1109/WIFS.2014.7084300.\\

\bibitem{phadikar2012novel}A. Phadikar, S.P. Maity, M. Mandal, Novel wavelet-based QIM data hiding technique for tamper detection and correction of digital images, J. Vis. Commun. Image Represent. 23 (2012) 454–466. doi:10.1016/j.jvcir.2012.01.005.\\

\bibitem{benrhouma2015tamper}O. Benrhouma, H. Hermassi, S. Belghith, Tamper detection and self-recovery scheme by DWT watermarking, Nonlinear Dyn. 79 (2015) 1817–1833. doi:10.1007/s11071-014-1777-3.\\

\bibitem{lin2011roi}S.D. Lin, J. Lin, C. Chen, A ROI-based semi-fragile watermarking for image tamper detection and recovery, Int. J. Innov. Comput. Inf. Control. 7 (2011) 6875–6888.\\

\bibitem{li2016semi}C. Li, R. Yang, Z. Liu, J. Li, Z. Guo, Semi-fragile self-recoverable watermarking scheme for face image protection, Comput. Electr. Eng. (2016).\\

\bibitem{cruz2015face}C. Cruz-Ramos, M. Nakano-Miyatake, H. Perez-Meana, R. Reyes-Reyes, L. Rosales-Roldan, Face region authentication and recovery system based on SPIHT and watermarking, Multimed. Tools Appl. 74 (2014) 7685–7709. doi:10.1007/s11042-014-2006-2.
\\

\bibitem{USC}The USC-SIPI Image Database, (2016). http://sipi.usc.edu/database/.\\

\end{thebibliography}

\end{document}